\documentclass[]{iopart}
\newcommand{\Conn}[3]{\Gamma_{#1}{}^{#2}{}_{#3}}

\usepackage[final]{graphicx}  % Include figure files
\usepackage{amssymb} % For proper R, C, Z fonts
\usepackage{epstopdf}
\usepackage{color}
\usepackage{url}
\usepackage{psfrag}

\begin{document}

%%%%%%%%%%%%%%%%%%%%%%%%%%%%%%%%%%%%%%%%%%%%%%%%%%%%%%%%%%%%%%%%%%%%%%
\title{Implementation of standard testbeds for numerical relativity}

\author{M~C~Babiuc$^1$, S~Husa$^{2,8}$, D~Alic$^3$, I~Hinder$^4$,
C~Lechner$^5$, E~Schnetter$^{6,7}$, B~Szil\'{a}gyi$^8$, Y~Zlochower$^9$,
N~Dorband$^8$, D~Pollney$^8$ and J~Winicour$^{8,10}$}
%\collaboration{Apples with Apples Alliance}

\address{1 Department of Physics and Physical Science, 
Marshall University, Huntington, WV 25755, USA}
\address{2 Friedrich Schiller University Jena, Max-Wien-Platz 1, 07743 Jena, Germany}
\address{3 Department of Physics, University of the Balearic Islands, Cra.~Valldemossa km.~7.5, 07122 Palma de Mallorca, Spain}
\address{4 Center for Gravitational Wave Physics, The Pennsylvania State University, 
University Park, PA 16802, USA}
\address{5 Weierstrass Institute for Applied Analysis and
Stochastics (WIAS), Mohrenstra\ss e 39, 10117 Berlin, Germany}
\address{6 Center for Computation \& Technology, 216 Johnston Hall,
  Louisiana State University, Baton Rouge, LA 70803, USA}
\address{7 Department of Physics and Astronomy, 202 Nicholson Hall,
  Louisiana State University, Baton Rouge, LA 70803, USA}
\address{8 Max-Planck-Institut f{\"u}r Gravitationsphysik
  (Albert-Einstein-Institut), Am M{\"u}hlenberg 1, 14076 Golm, Germany}
\address{9 Center for Computational Relativity and Gravitation,
   School of Mathematical Sciences,
   Rochester Institute of Technology, 78 Lomb Memorial Drive, Rochester,
   New York 14623, USA}
\address{10 Department of Physics and Astronomy,
University of Pittsburgh, Pittsburgh, Pennsylvania 15260, USA}

\begin{abstract}

We discuss results that have been obtained from the implementation of the
initial round of testbeds for numerical relativity which was proposed in the
first paper of the Apples with Apples Alliance. We present benchmark results for
various codes which provide templates for analyzing the testbeds and to draw
conclusions about various features of the codes. This allows us to sharpen the
initial test specifications, design a new test and add theoretical insight.

\end{abstract}

\pacs{04.70.Bw, 04.25.Dm, 04.40.Nr, 98.80.Cq}
%\maketitle

%%%%%%%%%%%%%%%%%%%%%%%%%%%%%%%%%%%%%%%%%%%%%%%%%%
\section{Introduction}
%%%%%%%%%%%%%%%%%%%%%%%%%%%%%%%%%%%%%%%%%%%%%%%%%%

For decades, the field of numerical relativity has been dominated by an often
painful quest for stable black-hole inspiral simulations. More than forty years
after Hahn and Lindquist's first pioneering numerical simulation of colliding
black holes~\cite{Hahn64}, this quest has recently turned into a gold-rush when
Pretorius's breakthrough simulation \cite{Pretorius:2005gq} based on a harmonic
code  was followed by  simultaneous invention of the ``moving punctures'' method
by two independent groups \cite{Campanelli:2005dd,Baker:2005vv}.

The primary motivation for solving the binary black hole problem in numerical
relativity has however been to supply waveforms for gravitational wave
detectors. This goal demands an approach that goes beyond the efforts that have
lead to an explosion in publications from the binary black hole community.
Cross-validation of waveforms between different groups (and codes) and
comparison with post-Newtonian predictions will be essential for numerical
waveforms to be used in the computationally expensive searches conducted by the
international gravitational wave community.  The importance of cross-validation
of numerical relativity results as a community effort was  foreseen by the
Apples with Apples Alliance (AwA) \cite{applesweb}, which has presented a first
round of standardized testbeds \cite{Alcubierre2003:mexico-I}. This first round
comprises four tests with periodic boundaries, designed to efficiently exhibit
code instability and inaccuracy. Instabilities currently receive less attention,
since it has turned out that, paradoxically, binary black hole evolutions are in
some sense a simpler problem than had been expected, and current codes evolving
binary black holes do not typically  show instabilities. The same codes {\em
will} however have difficulties with some of the testbeds presented in the first
round. The theoretical understanding of what works and what does not in
numerical relativity is still very much an open problem. One crucial theoretical
advance, which has been made since the publication of  our first
paper~\cite{Alcubierre2003:mexico-I}, is the development of a theory for
well-posed second order in space, first order in time systems
\cite{Beyer:2004sv,Gundlach04a,Gundlach:2004jp,Gundlach:2005ta,Calabrese:2005ft,
Szilagyi05,Motamed06}, which has been extended to a basic understanding of
numerical stability for such systems
\cite{Calabrese:2005ft,Szilagyi05,Motamed06}.

Over the past years several groups have committed their test results to a
publicly available data repository, with activities being coordinated via the
web-site  \url{http://www.ApplesWithApples.org}. The purpose of the present
paper is to document these developments and discuss their feedback with respect
to code performance, to test improvement and to design further tests.
While predating the binary black hole breakthroughs, we believe that the initial
Apples with Apples tests and results are still valuable as providing a first
testbed for a community effort in  numerical relativity.

The tests side-step many issues that would arise in a precise  discussion of the
binary black hole problem, such as the issue of boundaries. We make the natural
choice of periodic boundaries for a first round of tests to isolate the performance
of evolution algorithms. This is equivalent to evolution on the topology of a
3-torus in the absence of boundaries. However, in the context of general
relativity, this introduces complications of a cosmological nature regarding the
instability of Minkowski spacetime to perturbations on a compact manifold, as has
been discussed in~\cite{Alcubierre2003:mexico-I}.

Establishing a paradigm for standardized testbeds for numerical relativity is a 
formidable task in itself. We can draw on experience from other fields, such as
computational hydrodynamics where such testbeds have been used for a long time (for
an overview of CFD testbed resources on the web, see e.g.\
\cite{cfd-online-web-tests}; for an example of initial value ordinary differential
equation  (ODE) test-suites see \cite{ODEIVPweb}). However, general relativity comes
with its own issues that introduce extra  complications. First of all, it is
important to realize that the numerical relativity community is small, with very
limited available manpower. In contrast to the size of the field, we are trying to
solve many difficult problems at the same time. Numerical methods are being
developed in parallel with the formulation of the continuum problem, with the
construction of physically relevant initial data sets and with the unraveling of
the physical processes  involved in the systems under investigation. All of this
is, so far, without the help of comparison with experiments. Groups working in the
field are faced with many fundamental questions in designing their approaches.
Codes are in a state of flux that makes careful documentation easy to postpone. A 
good example is the issue of boundaries, which can be taken to be either a cubic
grid boundary or a smooth spherical boundary, which can either be mapped to
infinity or given some finite artificial location, and which are further
complicated by gauge freedom and the requirements of constraint preservation.
Useful comparison of the wide variety of resulting codes requires simple tests
which isolate an important facet of the problem.

We distinguish two fundamentally different types of testbed: The first type
compares different codes and methods in the treatment of a physically
interesting set of solutions. In the context of the binary black hole problem, a
detailed comparison of  nonspinning equal-mass inspiral would be a natural
example.  The second type are idealized situations, such as the  ``shock tube
test'' \cite{Courant62b} in computational fluid dynamics. This is the type of
testbed we discuss in the present paper,
where we restrict ourselves to a greatly simplified first set of tests
\cite{Alcubierre2003:mexico-I}: periodic grids and strict test specifications,
which as far as practicable define all the details of a simulation except the
formulation of the Einstein equations. Our experience with the first round of
testbeds confirms this decision: even the analysis of these simple situations
has proved quite challenging. Our conclusions in Sec.~\ref{sec:discussion}
discuss how the experience from the present round of tests can be used in our
development of  black hole tests. 

We identify five main aims of standardized tests of the ``idealized'' type:
\begin{enumerate}
\item Standardized tests should provide the young and fast-changing community
of numerical relativists with a common reference frame which will help
integrate different efforts to produce a coherent picture of what works and
what does not, and thus reduce the dependence on anecdote and fashion.
\item Tests should be efficient in revealing instabilities or other
weaknesses of an algorithm, both regarding simplicity of the analysis, run
time and implementation.
\item Tests should help identify where problems come from, as a step toward
improvement of the algorithms.
\item Tests should facilitate comparisons between approaches
regarding different continuum formulations, spatial discretizations, time
integrators, uses of artificial dissipation, etc.
\item The development of testbeds should eventually lead to useful code
comparisons for judging the validity of  physically interesting simulations,
e.g.\ the binary black hole problem.
\end{enumerate}

Point (i) has been addressed by organizing this project as a community
initiative, which seeks broad participation and provides test results via web
pages and a CVS repository~\cite{applesweb}. Regarding point (ii), in this paper
we review our original test specifications and propose modifications to promote
efficiency. Point (iii) is essential for the character of this paper: we focus
on presenting test results as a template for analyzing and interpreting
results, rather than just presenting the broadest possible listing of test
output for a maximal number of codes. We feel that it is essential to stress
this point: tests which do not directly correspond to a physically interesting
situation are only valuable if they improve our understanding of what really
goes on with a certain code. Only then can we hope to carry over test benefits
to other situations. Such analysis does of course require a certain effort.

Point (iv) is dealt with by providing ``standard candle results'' in the CVS
repository, i.e., benchmarks  that have been obtained with very strictly defined
specifications. Point (v) represents the ultimate goal of the AwA Alliance.

The analysis of test results has led to better understanding of the four
original standardized tests and has led to some improvements in their
specifications. We also have added a new shifted gauge wave test, which closes a
gap regarding the ability of a code to handle a shift. The revised
specifications for the five tests are detailed in \ref{sec:revspecs}. The major
changes from the specifications in \cite{Alcubierre2003:mexico-I} are

\begin{itemize}

\item {\bf Robust stability test:} The rules for how the data should scale
with resolution have been changed; the criteria for passing the test.
has been restated.

\item {\bf Linearized wave test:} No changes.

\item {\bf Gauge wave test:} The original tests amplitudes $A=.01$ and
$A=.1$ have been replaced with $A=.5$.

\item {\bf Shifted gauge wave test:} This new test has been added.

\item {\bf Gowdy wave test:} No changes.

\end{itemize}

We have also dropped the original requiremnt that the tests be run with
a iterative Crank-Nicholson integrator. 
Conclusions from the test results and our experiences with the testing 
procedures, along with the reasons behind the changes and additions in the
standard tests, are summarized in Sec.~\ref{sec:discussion}.

The code descriptions and test data on which this paper is based are described
in Sec.~\ref{sec:formulations}. The results for the original four standardized
tests are discussed in Secs. \ref{sec:robusttest},  \ref{sec:linwavetest},
\ref{sec:gaugewavetest} and \ref{sec:gowdywavetest}. Discussion of the shifted
gauge wave test and some benchmarks are given in Sec.~\ref{sec:shiftedtest}. 

The plots presented in this paper are based upon test output in the CVS
repository. Many of these tests were run with codes in which artificial
dissipation was only introduced implicitly through the use an iterated
Crank-Nicholson (ICN) time integrator. It had been a naive hope at the beginning
of this project that the use of ICN might provide a way to standardize the
introduction of dissipation. Most numerical relativity groups now use
Runge-Kutta time integrators with the explicit addition of Kreiss-Oliger
dissipation (see \ref{sec:appendix_dissipation}). It has been found that many of
the test results presented here could be greatly improved by such explicit use
of dissipation. In addition to artificial dissipation, most codes that
simulate binary black holes use higher order approximations than the second
order accurate codes being compared here. Consequently, we want to emphasize
that the results exhibited in this paper should not be used to make judgments on
particular approaches, but that our purpose is to assess and improve the test
suite and to provide a basis for future code comparisons.

%%%%%%%%%%%%%%%%%%%%%%%%%%%%%%%%%%%%%%%%%%%%%%%%%%
\section{Code descriptions}
\label{sec:formulations}
%%%%%%%%%%%%%%%%%%%%%%%%%%%%%%%%%%%%%%%%%%%%%%%%%%

In order to ensure a consistent presentation of test output, we present a brief account
of the numerical codes and algorithms which have been used to produce the data on which
this paper is based. All data are publicly available via the CVS repository
(see~\cite{applesweb} for details). The four original standardized tests are denoted by
ROBUST (the Robust Stability Test), LINEAR (the Linear Wave Test), GAUGE (the Gauge
Wave Test) and GOWDY (the Gowdy Wave Test). Table 1 summarizes the output data that
have been  submitted for the various codes.

\begin{table}[htbp]
  \begin{center}
    \begin{tabular}[c]{|c|c|c|c|c|}
\hline
  CODE &
   ROBUST  &
   LINEAR   &
   GAUGE  &
   GOWDY
    \\ \hline \hline
%%%%%
Abigel\_harm                 & $++$ & $++$ & $++$  & $++$
\\ \hline
%%%%%
%%%%%
AEI\_CactusEinsteinADM          & $+$ & $--$ & $--$  & $++$
\\ \hline
%%%%%
%%%%%
Kranc\_FreeADM         & $+$ & $+$ & $+$ & $+$
\\ \hline
%%%%%
CCATIE\_BSSN          & $++$ & $++$ & $++$ & $++$
\\ \hline
%%%%%
%%%%%
Kranc\_BSSN             & $++$ & $++$ & $++$ & $++$
\\ \hline
%%%%%
%%%%%
LazEv\_BSSN             & $++$ & $++$ & $++$ & $++$
\\ \hline
%%%%%
%%%%%
HarmNaive              & $++$ & $++$ & $++$ & $++$
\\ \hline
%%%%%
%%%%%
KrancNOR               & $++$ & $++$ & $++$ & $++$
\\ \hline
%%%%%
%%%%%
KrancFN                & $++$ & $++$ & $++$ & $--$
\\ \hline
%%%%%
%%%%%
LSU\_HyperGR            & $++$ & $++$ & $++$ & $++$
\\ \hline
%%%%%
\end{tabular}
    \caption{Test output and codes considered in this article.
      The code abbreviations are explained below, along with a
      description of the finite difference algorithm.
       A ``$++$'' indicates a full complement of test output in the CVS,
a ``$+$´´ indicates partial output which has been used for our analysis,
a ``$-$'' indicates partial output on which no meaningful conclusions could be
drawn and a ``$--$'' indicates no output.}
    \label{tab:cvs}
  \end{center}
\end{table}

The usefulness of this data depends upon good code documentation. It is beyond
the scope of this paper to provide such documentation for all the codes
involved. However, we will outline
some basic code information which is necessary to interpret the test results.
The complexity of this task is somewhat alleviated because all the   codes
represented here follow a method of lines approach. We will organize the code
descriptions along the  following guidelines.

\begin{itemize}
\item A description of the {\em continuum formulation},
	including a list of all variables, their associated evolution
	equations and constraints (both differential and algebraic),
	equations governing the lapse and shift and a specification
	of any free parameters.
	Terms and differential operators in the equations should be
	ordered in the way that they are approximated by finite difference
	expressions in order to avoid ambiguities associated with the
	Leibniz rule. The hyperbolicity classification should be provided, if
	known.

\item A description of the {\em semi-discrete system}, describing
      the spatial finite difference equations on each time level, including
      the rules for discretizing partial derivatives
      as centered or one-sided finite differences and
      any other discretization techniques, such as spatial averaging or
      dissipation. For complicated systems, the finite difference rules may
      be specified only for the principal part,
      with further details supplied by references.
      (Here we provide some basic reference material
      in~\ref{sec:appendix_codes} and~\ref{sec:appendix_numerics}
      for compactness of presentation.)

\item A description of the {\em numerical time update scheme}.
	All manipulations of data between intermediate time
	steps should be specified, such as enforcing a constraint.
\end{itemize}

As an example, we consider two inequivalent algorithms for
the wave equation $\Box \phi = 0$ (with unit lapse, zero shift
and spatial metric $\gamma_{ij}$),
which should be expected to result in different code performance.
In both cases the second order in time system is reduced to first order in
time by introducing the variable
$
    \pi=\partial_t \phi,
$
and applying, say, 4th order Runge-Kutta (see \ref{sec:appendix_numerics}) to
the ODEs of the semi-discrete  system obtained using the method of lines. Two
different codes can based upon the following descriptions.

Description {\bf I}:
\begin{enumerate}
\item
The continuum system is
\begin{eqnarray}
       \partial_t \phi &=& \pi, \\
\partial_t \pi &=&
\frac{1}{\sqrt{\gamma}}\partial_i(\sqrt{\gamma}\gamma^{ij}\partial_j \phi).
\label{eq:wave1}
\end{eqnarray}
\item
The semi-discrete version is obtained by replacing all partial
derivatives in (\ref{eq:wave1}) by centered differences:
$$
       \partial_t \pi = \frac{1}{\sqrt{\gamma}} 
        D_{0i}(\sqrt{\gamma} \gamma^{ij}D_{0j} \phi),
$$
where $D_{0i}$ is the centered difference
operator $D_0$ applied in direction $i$ (see~\ref{sec:appendix_FD}).
\end{enumerate}

Description {\bf II} (inequivalent with {\bf I}):
\begin{enumerate}
\item The continuum system is
\begin{eqnarray}
     \partial_t \phi &=& \pi, \\
     \partial_t \pi &=& \gamma^{ij}\partial_i \partial_j \phi 
      + \frac{1}{\sqrt{\gamma}}
          \partial_i(\sqrt{\gamma} \gamma^{ij}) \partial_j \phi.
\end{eqnarray}
\item
The semi-discrete version is obtained by replacing the partial
derivatives in (\ref{eq:wave1}) by centered differences according to
\begin{eqnarray}
    \gamma^{ij}\partial_i \partial_j \phi + \frac{1}{\sqrt{\gamma}}
          \partial_i(\sqrt{\gamma} \gamma^{ij}) \partial_j \phi
            \nonumber \\
          = \gamma^{ij}D_{+i}D_{-j} \phi + \frac{1}{\sqrt{\gamma}}
            D_{0i}(\sqrt{\gamma} \gamma^{ij})  D_{0j} \phi
\end{eqnarray}
where $D_{+i}$ and $D_{-i}$  represent forward and backward
centered finite differences in the respective directions
(see~\ref{sec:appendix_FD}).
\end{enumerate}

The codes resulting from these two descriptions
produce substantially different performance because of
the ``checkerboard'' design of the stencil used in description {\bf I}.
Descriptions of the specific codes used in this paper
are given in~\ref{sec:appendix_codes}.

%%%%%%%%%%%%%%%%%%%%%%%%%%%%%%%%%%%%%%%%%%%%%%%%%%
\section{Robust stability test}\label{sec:robusttest}
%%%%%%%%%%%%%%%%%%%%%%%%%%%%%%%%%%%%%%%%%%%%%%%%%%

The robust stability test was intended as a first screen to eliminate many
unstable evolution algorithms. The particular importance of this test was due to
the fact that instabilities of numerical codes appeared as a prime obstacle to
``solve'' the binary black hole problem, and essentially no theoretical
understanding was available to discuss the well-posedness and numerical
stability of first order in time, second order in space formulations of the
Einstein equations, which have been and still are popular in the field.
Recently, a theoretical framework has become available to discuss the
well-posedness and numerical stability of such mixed order formulations of the
Einstein equations
\cite{Nagy:2004td,Beyer:2004sv,Gundlach04a,Gundlach:2004jp,Gundlach:2005ta,
Gundlach:2006tw,Calabrese:2005ft,Szilagyi05,Motamed06},
and it has been extended to the problem of discretizing the equations in the
context of the method of lines \cite{Calabrese:2005ft,Szilagyi05,Motamed06}. As
a consequence of both the recent breakthroughs in the binary black hole problem
and the theoretical advances, numerical stability has become a {\em relatively}
minor issue in practice (although there certainly remain interesting
mathematical questions to be pursued). We thus restrict ourselves to a minimal
discussion here, as is sufficient to understand the data available in our test
results repository. For a more in-depth discussion of theoretical and practical
aspects of numerical stability and  the robust stability test we refer to
\cite{Calabrese:2005ft}, which has been directly motivated by numerical results
obtained within this project.

While the other tests give quantitative
information about an evolution system, e.g.\ the magnitude of the numerical
error, the  result of the robust stability test is ``pass'' or ``fail''. A
stable numerical algorithm is only possible if the underlying continuum problem
is well-posed \cite{Calabrese02a}. 
In the well-posed case an instability might still
arise, either from the numerical technique or from the existence of an
exponential mode in the continuum problem. The test is designed to avoid
continuum instabilities by considering small perturbations of the Minkowski
metric. In addition to providing efficient detection of unstable numerical
algorithms (or coding errors) affecting the principal part of the evolution
system, it is also intended to spot instabilities arising from ill-posed
systems, such as weakly hyperbolic systems.

As an example, consider the weakly hyperbolic system
\begin{eqnarray}
      u_{,t}&=&u_{,x}+v_{,x} \nonumber\\
      v_{,t}&=&v_{,x}
      \label{eq:weakeq}
\end{eqnarray}
with the periodic solutions
\begin{eqnarray}
    u &=& \omega t\cos\omega(t+ x) \, , \quad v=\sin\omega(t+ x) \,
                   \label{eq:weaksol}        \\
            \omega &=& 2\pi m \, , \quad m=1,2,3,... \nonumber 
\end{eqnarray}
on the domain $-.5 \le x \le .5$.
In terms of the $L_2$ norm
\begin{equation}
      N=\bigg (\int_{-.5}^{.5} (u^2+v^2)dx \bigg )^{1/2},
\end{equation}
the Cauchy data for (\ref{eq:weaksol}) at $t=0$,
\begin{equation}
    u=0 \, , \quad v=\sin\omega x,
    \label{eq:sol}
\end{equation}
has norm $N(0)=1/\sqrt{2}$.
However, because of (\ref{eq:weaksol}), $N(t)\sim \omega t$
for large $\omega$.
This leads to a violation of the well-posedness requirement
that in any finite time interval
\begin{equation}
      N(t)< Ae^{Kt} N(0),
\end{equation}
in terms of constants $A$ and $K$ independent of the Cauchy data. 

For discretized systems we can not test well-posedness directly, but
rather we test the analogous concept of numerical stability, i.e., we
aim at establishing the existence of constants $A$ and $K$,
which give rise to the bound
\begin{equation}
\frac{\| v^n \|}{\| v^0 \|} \le Ae^{K t_n},\label{def:stability}
\end{equation}
where $v^n$ is the solution of the discrete system at time $t_n=nk$. The test is
passed if such a bound can be established, and is failed otherwise. In the
discretized version of a weakly hyperbolic problem, with grid displacement
$h$, the perturbation of a simulation by random initial data can be expected to
excite numerical error which grows linearly in time according to $u\sim t/h$,
corresponding to the shortest wave number $\omega\sim 1/h$. This would then lead
to secular error growth which increases with resolution. Although the system
(\ref{eq:weakeq}) is well-posed with respect to a stronger norm including a
$v_{,x}^2$ term, a generic perturbation of (\ref{eq:weakeq}) by lower order
terms would nevertheless produce an exponentially growing instability which
cannot be bounded. See~\cite{Gustafsson95} for a more general discussion of
such weakly hyperbolic systems.

The key idea of setting initial data for this test is to distribute energy roughly
equally over all frequencies. This is a particularly efficient way to reveal growing
modes if the growth rate increases with resolution, as is the  case if the
discretization is unstable or if the continuum problem is ill-posed. In our test we use
a spectrum generated by random initial data.

The robust stability test as formulated here tests numerical stability in the
linear, constant coefficient regime. It is based upon small random perturbations
of Minkowski space, with the initial data consisting of random numbers
$\epsilon$ applied at each grid point to every code variable requiring
initialization.  In numerical evolution, where machine precision takes the place
of $\epsilon$, a code that cannot stably evolve such random noise would be
unable to evolve smooth initial data.

In spite of its simplicity, our experience has shown that the robust stability
test  exhibits various subtle difficulties  in designing a single test
prescription that is universally effective  for all evolution systems and
numerical methods. Some particular problems are:

\begin{itemize}

\item For random initial data, where a significant part of the  total energy is
in high frequencies, dissipation has a large effect. Some intrinsic dissipation
is unavoidable in finite difference evolution algorithms,  and adding artificial
dissipation may be necessary to stabilize certain  algorithms
\cite{Calabrese:2005ft}, and insufficient to stabilize others (such as
algorithms for weakly hyperbolic systems). Simulations of variable coefficient,
nonlinear systems normally require numerical dissipation
to obtain a stable evolution, e.g.\ by adding Kreiss-Oliger type
dissipation~\cite{Gustafsson95} (see~\ref{sec:appendix_dissipation}). Dissipation
can however increase the time scale on which instabilities become apparent. The
detailed way dissipation affects instabilities varies with the
spatial discretization (we only consider second order approximations
here), with the time integrator, with the grid resolution and with the Courant
number.

\item
As discussed in the above example, well-posedness and numerical stability are
defined with respect to a certain norm. Using
an inappropriate norm can yield misleading results. Second order systems
require different norms than first order systems \cite{Calabrese:2005ft}.

\item Numerical stability of an explicit time integration algorithm can only be
expected if the time step is appropriately restricted by a
Courant-Friedrichs-Lewy (CFL) condition. It is important to distinguish between
resolution dependent blowup associated with ill-posedness from blowup resulting
from a CFL violation. For sufficiently complicated 3D algorithms, the CFL limit
might not be readily deduced from analytic arguments. As an example, exponential
growth of the ADM algorithm was mistakenly provided as an illustration of a
failed robust stability test in~\cite{Alcubierre2003:mexico-I} . It took
subsequent testing and analysis to reveal that this exponential growth resulted
from a CFL violation and that otherwise the weakly hyperbolic instability of ADM
resulted in a secular (linear in time) growth.

\end{itemize}

As  a result of such considerations, we will not
try to present a single universally applicable specification
for the robust stability test. Instead, while keeping the original spirit
of the test as a simple and useful first screen, we propose some changes
in the guidelines, as discussed below.

An important issue when performing stability tests is whether the high frequency
modes are damped. This has important bearing on the long-time behavior of the
robust stability test: all damped modes will decay in time;  eventually the
undamped frequencies of the discrete system will dominate the signal. If an
analysis of damping factors has not been performed, the test can therefore also
be useful in detecting the spectrum of frequencies which are not damped. It has
been pointed out in~\cite{Calabrese:2005ft} for standard discretizations of
first order in space systems that the ``checkerboard'' mode is undamped, while
for typical second order systems it is damped. Since the ``checkerboard'' mode
is not realized on grids with an odd number of points, we adopt the practice of
always using an even number of grid points so as not to muzzle such a potential
instability.

In our original specifications, we proposed the relatively large time step $dt =
0.5 dx$, which turned out to be larger than the CFL limit for the ADM system.
Since a smaller $dt$ also decreases the amount of dissipation inherent in a time
integrator, we now propose a relatively small time step to avoid distortion of
results due to dissipation. Common time integrators in current practice in
numerical relativity are ICN, RK3 and RK4 (sorted by decreasing internal amount
of dissipation). A sufficiently small time step would yield similar results for
all of them. We therefore propose to run with $dt = 0.1 dx$, which can be
further reduced in case of doubt. See \ref{sec:rrobust} for details.

For systems that use variables which correspond to
spatial derivatives of the ADM 3-metric and extrinsic curvature, an ambiguity
arises: noise can be added uniformly to all variables, or to the ADM initial
data before taking derivatives. There are similar ambiguities in second order
systems regarding how the range of the random numbers should scale with
resolution. For uniformity of description, we propose to do the simplest thing,
namely to apply noise to all evolution variables in the same way. We propose
the range of $\pm 10^{-10}$ for all variables, the same range used for the
lowest resolution in the original specifications.

Following common practice at the time, the Hamiltonian constraint was used to
analyze test results. Again following~\cite{Calabrese:2005ft},  we now propose a
pass/fail analysis based upon whether the time behavior of the norm satisfies
(\ref{def:stability}).

Our core test specification combines both 1D and 3D features by running in a
thin channel along the $x$-axis. The use of 4 distinct gridpoints in the $y$ and
$z$ directions allows for the checkerboard mode (ghost points may be necessary
depending upon the numerical scheme). The generalization to a full cube 3D test
is straightforward, and may add further clarification in case of dubious
results. 

The test should be run until one is confident that dissipation effects do not
cloud the result. Without artificial dissipation, a runtime of one crossing
time, using output at every time step, is usually sufficient. This corresponds
to $500 \rho$ time steps, for a given resolution $\rho$ (see
\ref{sec:revspecs}). The test is passed if the norm satisfies the inequality
(\ref{def:stability}) for all resolutions, for a fixed choice of $A$ and $K$.

\begin{figure}
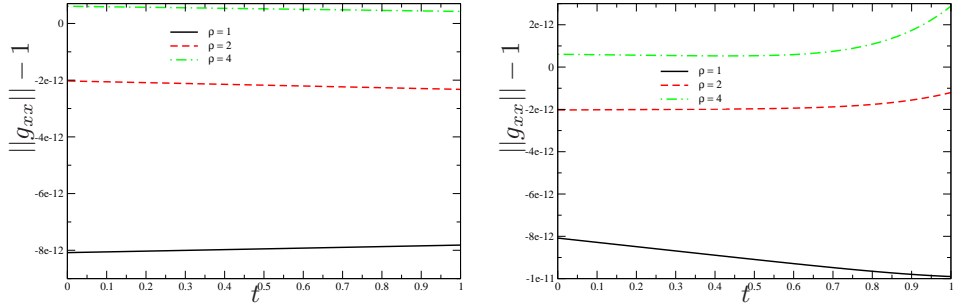

\begin{center}
  \psfrag{xlabel}{$t$}
  \psfrag{ylabel}{$||g_{xx}||-1$}
  \includegraphics*[width=14pc]{mex1resfig01}\hspace{5mm}
  \includegraphics*[width=14pc]{mex1resfig02}
\end{center}
  \caption{Convergence results for the robust stability test with 
the Abigel\_harm (left) and HarmNaive (right) codes, for runs of $1$  crossing
time. The graphs show the error in $g_{xx}$ as a function  of time, obtained by
subtracting 1 from its $L_2$ norm. As seen from the slopes of the graphs, the 
Abigel\_harm code (left) passes the test, because there is no increasing rate of
error growth with higher resolution $\rho$, while the HarmNaive code (right)
fails the test because the growth rate increases with resolution.
}
\label{fig:RobustHarm.gxxnorm2}
\end{figure}

Instabilities caused by the ill-posedness of the evolution system (or by coding
errors in treating the principal part), are already apparent in one-dimensional
tests, which can be performed quickly and economically. An example of how this
analysis works  is given in Fig.~\ref{fig:RobustHarm.gxxnorm2}. The way that the
slope of the error {\it vs} time depends upon  resolution shows that the
Abigel\_harm code, which  is based upon a symmetric hyperbolic formulation,
passes the test; whereas the  HarmNaive code, which is based upon a weakly
hyperbolic formulation, fails the  test. 

%

%%%%%%%%%%%%%%%%%%%%%%%%%%%%%%%%%%%%%%%%%%%%%%%%%%
\section{Linearized wave test}\label{sec:linwavetest}
%%%%%%%%%%%%%%%%%%%%%%%%%%%%%%%%%%%%%%%%%%%%%%%%%%

A prime physical objective of numerical relativity is to compute the waveform
from a system of black holes and neutron stars. This test checks the ability of
a code to propagate a linearized gravitational wave, which is a minimally 
necessary attribute for reliable wave extraction from strong sources. 
Test specifications are given in \ref{sec:rlinwave}.

The test checks the accuracy of the code in propagating both the amplitude and
phase of the wave. It can reveal whether excessive dissipation has been
necessary for good long term performance in the robust stability
test. For the $\rho =1$ coarsest grid ($N=50$ grid zones), there is not enough
resolution for second order accurate codes to obtain accurate phase propagation
and the corresponding runs should only be viewed as an economical
first check on the code. The most useful comparisons are with the $\rho=4$
grid.

\begin{figure}[hbtp]
\begin{center}
  \psfrag{xlabel}{$x$}
  \psfrag{ylabel}{$g_{yy}-1$}
  \includegraphics*[width=20pc]{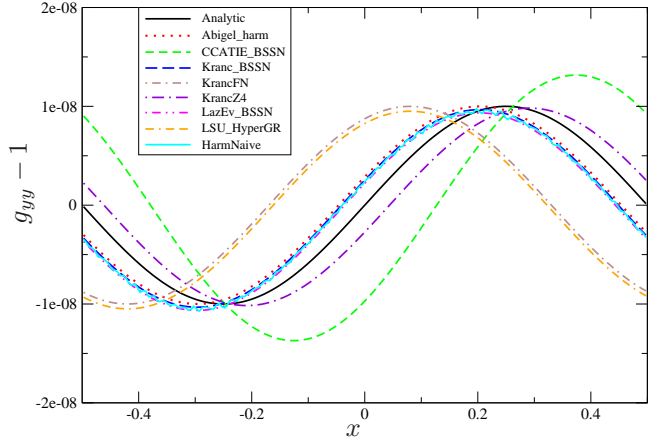}
\end{center}
  \caption{Comparison snapshots of $g_{yy}(x)-1$ at $t=1000$ for the 1D
     linearized wave test, with $\rho = 4$ resolution.}
\label{fig:LPWave1Dcomp4gyym1}
\end{figure}

\begin{figure}[hbtp]
\begin{center}
  \psfrag{xlabel}{$x$}
  \psfrag{ylabel}{$g^{err}_{yy}$}
  \includegraphics*[width=20pc]{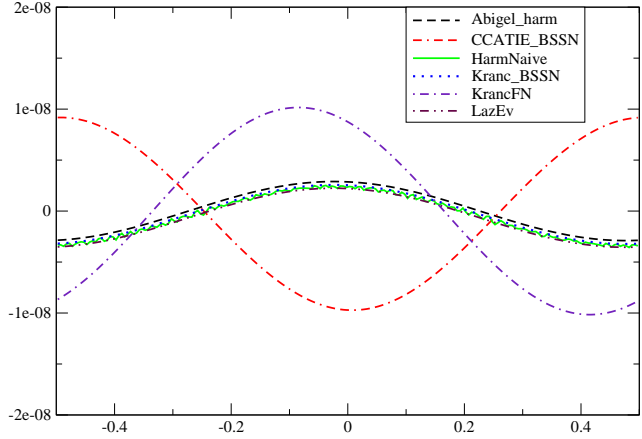}
\end{center}
  \caption{Comparison snapshots of the error ${\cal E}$ in $g_{yy}(x)$ at $t=1000$ 
           for the 1D linearized wave test, with $\rho = 4$ resolution.}
\label{fig:LPWave1Dcomp4Egyy}
\end{figure}

Fig.~\ref{fig:LPWave1Dcomp4gyym1} compares snapshots of the 1D wave after 
$1000$ crossing times which were obtained with a variety of codes using the
$\rho=4$ finest grid. For reference, the exact waveform is also plotted. The
snapshots for three of the codes, Abigel\_harm, HarmNaive and LazEv\_BSSN, are
very similar and provide a good benchmark for the accuracy that can be achieved
at this resolution. They very closely match the exact solution in amplitude  but
show a phase delay, similar to the delay seen in the following gauge wave test.
It should be expected that phase accuracy could be improved by going to fourth
order accurate methods. Some snapshots of the corresponding error are displayed
in Fig.~\ref{fig:LPWave1Dcomp4Egyy}. Except for the two codes with the largest
phase error, the error at $1000$ crossing times is confined to a small band. By
monitoring the growth of the error during the evolution, it was verified that no
overall multiple of $2\pi$ phase error is concealed in the snapshots of
Fig.~\ref{fig:LPWave1Dcomp4gyym1}.


\begin{figure}[hbtp]
\begin{center}
  \psfrag{xlabel}{$t$}
  \psfrag{ylabel}{$||{\cal H}||$}
  \includegraphics*[width=20pc]{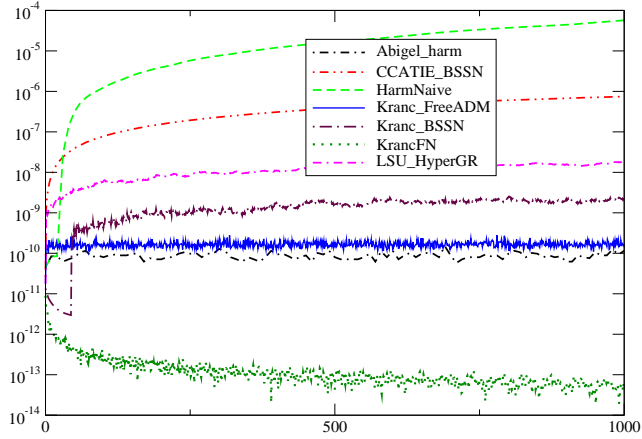}
\end{center}
  \caption{Comparison of time dependence of the $L_\infty$ norm of the
Hamiltonian constraint $||{\cal H}||$, shown on a logarithmic 
scale, for the 1D linearized wave test with $\rho = 4$ resolution.
}
\label{fig:LinWr4loghamnorminfcomp}
\end{figure}

In addition, the plots of the Hamiltonian in Fig.~\ref{fig:LinWr4loghamnorminfcomp} show no rapidly growing
constraint violating instabilities in this linear regime. The secular
instability of Harm\_Naive, which was discussed in the robust stability test, is
evident but it does not introduce a large error in this test. This illustrates
that instabilities associated with a weakly hyperbolic system are not
necessarily evident in linearized tests where, as discussed in
Sec.~\ref{sec:robusttest}, the unstable modes only grow secularly in time. The
KrancFN code gives good accuracy for the amplitude but a much larger error in
phase. The CCATIE code shows poor accuracy in both phase and error. It is beyond
the scope of this paper to explain the discrepancy between the performance of
the two BSSN codes.

The 1D linear wave test is simple and economical to perform. Although the test
is not very demanding, the results for the metric component $g_{yy}$ in
Figs.~\ref{fig:LPWave1Dcomp4gyym1} and \ref{fig:LPWave1Dcomp4Egyy} show that it
provides a benchmark which can be useful to identify weaknesses in code
performance. The 2D tests require more computer time and the results were
typically in line with expectations from the 1D results. 

%

%%%%%%%%%%%%%%%%%%%%%%%%%%%%%%%%%%%%%%%%%%%%%%%%%%
\section{Gauge wave test}\label{sec:gaugewavetest}
%%%%%%%%%%%%%%%%%%%%%%%%%%%%%%%%%%%%%%%%%%%%%%%%%%

The gauge wave test is based on a nonlinear gauge transformation of Minkowski
spacetime. Although the correct solution is a flat spacetime, nonlinear effects
and the nontrivial geometry of the time slices can easily trigger continuum
instabilities in the equations.  For simple examples of such effects see
\cite{Babiuc:2004pi} for a nonlinear wave equation on flat space, designed to
model problems arising in this testbed, and \cite{Husa:2005ns} for a linear
example of how nontrivial geometry of the slicing can trigger instabilities
already for the Maxwell equations.

Our original specifications~\cite{Alcubierre2003:mexico-I} were to run the
test with amplitudes $A=0.01$ and $A=0.1$. Many codes have been sufficiently
improved to handle larger amplitudes, which is generally more efficient in
detecting instabilities with smaller run times. Accordingly, we specify an
amplitude of $A=0.5$ in the revised test details given in \ref{sec:rgwave}.

While the gauge wave metric has a rather simple form, the test proved to be
challenging for most evolution codes. One anticipated source of growing error is
the instability of a flat space with $T^3$ topology
\cite{Alcubierre2003:mexico-I}. Another problem is the existence of a family of
harmonic, exponential gauge modes corresponding to the substitution $H
\rightarrow e^{\lambda t} H$ (for arbitrary $\lambda$) in the metric
(\ref{eq:gwmet})~\cite{Babiuc:2004pi}. The testbed itself corresponds to
$\lambda=0$, but numerical error can easily excite this mode and lead to
exponential growth of the wave amplitude. Other instabilities may be present in
individual systems, depending on the detailed form of the reduced evolution
system for the particular formulation. Some of these instabilities can be
identified by looking at the growth of the constraints for the formulation. In
addition to instabilities that correspond to solutions of the continuum problem,
individual codes may suffer from numerical instabilities depending on the
discretization schemes.  These would typically be  seen as high frequency modes
and, for well-posed systems, can be cured by adding artificial dissipation to
the numerical algorithm.

\begin{figure}[hbtp]
\begin{center}
  \psfrag{xlabel}{$t$}
  \psfrag{ylabel}{$||{\cal H}||$}
  \includegraphics*[width=20pc]{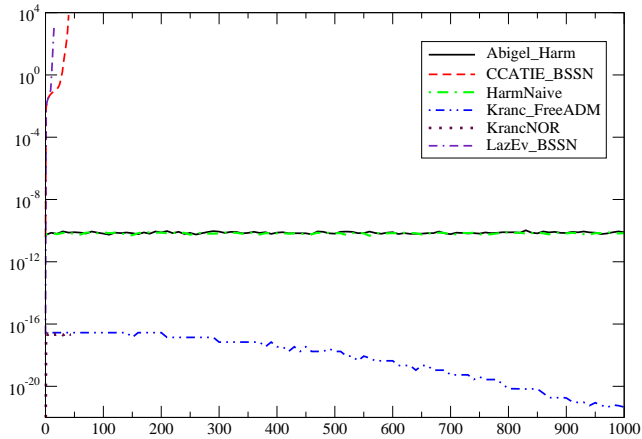}
\end{center}
  \caption{Time dependence of the $L_\infty$ norm of the Hamiltonian 
           constraint $||{\cal H}||$, shown on a  logarithmic scale, for the 
           1D gauge wave test with resolution $\rho =4$ and amplitude $A=0.1$.}
\label{fig:Gauge1Dhamcomplog}
\end{figure}

Figure~\ref{fig:Gauge1Dhamcomplog} shows the time evolution of the Hamiltonian
constraint for the various codes. The negligible violation of the Hamiltonian
constraint by the harmonic codes can be attributed to the fact that the harmonic
coordinate conditions are used to shift the role of the constraint to an
evolution equation. Note that the BSSN codes show rapid growth of Hamiltonian
constraint violation. So far no BSSN code has demonstrated satisfactory
performance for this test, and for brevity we do not include BSSN results in the
below results.

%

%%%%%%%%%%%%%%%%%%%%%%%%%%%%%%%%%%%%%%%%%%%%%%%%%%
\subsection{Results}
%%%%%%%%%%%%%%%%%%%%%%%%%%%%%%%%%%%%%%%%%%%%%%%%%%

%%%%%%%%%%%%%%%%%%%%%%%%%%%%%%%%%%%%%%%%%%%%%%%%%%
\subsubsection{Results for the Abigel\_harm Code}
%%%%%%%%%%%%%%%%%%%%%%%%%%%%%%%%%%%%%%%%%%%%%%%%%%

\begin{figure}[hbtp]
\begin{center}
  \psfrag{xlabel}{$x$}
  \psfrag{ylabel}{$\bar g^{zz}$}
  \includegraphics*[width=20pc]{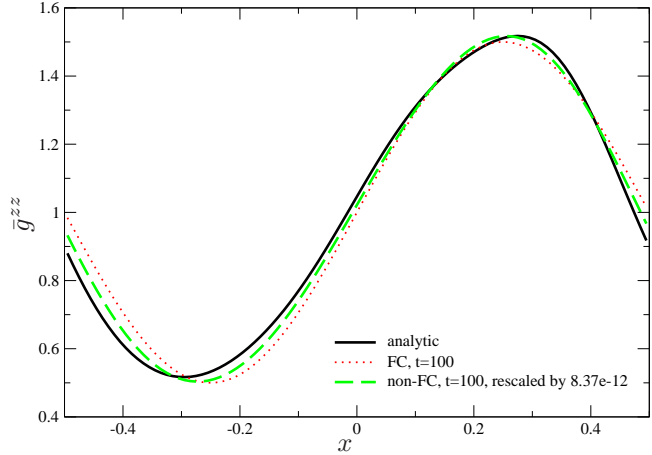}
\end{center}
\caption{Comparison of code performance between the non-flux-conservative
(non-FC) and flux-conservative (FC) versions of the Abigel\_harm code, showing
graphs of $\bar g^{zz}(x)$ at $t=100$ for a gauge wave of amplitude $A=0.5$ on
the $\rho=2$ grid. In the non-FC case the graph is rescaled by the average of
the plotted function, showing $\bar g^{zz} / avg(\bar g^{zz}) \approx \bar
g^{zz} / \exp(29.8)$.  The good overlap of this rescaled function with the
analytic value clearly indicates that the dominant error of the non-FC code is a
multiplicative function of $t$.  Measurements at $t=100$ for the non-FC code
show that logarithm of the spatial average of $\bar g^{zz}$ scales roughly as
$(dx)^2$, i.e., $\log(avg(\bar g^{zz})_{\rho=1}) \approx 110.8, \log(avg(\bar
g^{zz})_{\rho=2}) \approx 29.8, \log(avg(\bar g^{zz})_{\rho=4}) \approx 7.52$,
suggesting that the multiplicative error has exponential growth of the form
$\exp(O((dx)^2) \cdot t)$.
}
\label{fig:Abigel.nonFC.gammaxx}
\end{figure}

For this particular testbed most components of the densitized metric
$\bar g^{\mu\nu}=\sqrt{-g}g^{\mu\nu}$ have
trivial values, the non-trivial ones being
\begin{equation}
   \bar g^{yy} = \bar g^{zz} = H
\end{equation}
The original implementation of the Abigel code based upon
(\ref{eq:gaugewave.abigel.2nd}) leads to a numerically stable and convergent
code, with no high frequency modes generated. However, as shown by the dramatic
growth of the rescaled error plotted in Fig.~\ref{fig:Abigel.nonFC.gammaxx}, the
gauge wave excites exponential modes $\bar g^{yy} = \bar g^{zz}=1-e^{\lambda
t}H$, $\lambda > 0$. This can be understood~\cite{Babiuc:2004pi}in terms of
solutions of the harmonic system whose densitized metric components are all
trivial except for
\begin{equation}
       \bar g^{yy} = \bar g^{zz} = F(t,x).
\label{eq:gaugewave.abigel.gamma.F1D}
\end{equation}
The resulting source term $S^{\mu\nu}$ in (\ref{eq:gaugewave.abigel.2nd})
vanishes except for the components   \begin{equation}  S^{yy} = S^{zz} = \frac{
-F_t^2 + F_x^2 } {F}. \label{eq:gaugewave.abigel.src.F1D}   \end{equation}   The
PDE for $F(t,x)$, which results from inserting
(\ref{eq:gaugewave.abigel.gamma.F1D}) into (\ref{eq:gaugewave.abigel.2nd}),
reduces to $(-\partial_t^2 + \partial_x^2) \log F = 0$, which admits the
exponential solutions $F = e^{\lambda t} H$.  These solutions satisfy the
harmonic constraints and the reduced harmonic system
(\ref{eq:gaugewave.abigel.2nd}), so that they are also solutions of the full
Einstein equations.  Therefore all codes using harmonic gauge conditions might
be expected to excite this mode.

In the case of the Abigel\_harm code, these modes were suppressed by building
semi-discrete conservation laws into the code which, for the gauge wave initial
data, would not be obeyed by the exponential solution. Namely, by writing
(\ref{eq:gaugewave.abigel.2nd}) in the flux-conservative form
(\ref{eq:gaugewave.abigel.FC}), the principle part of the resulting equation has
vanishing source term, $\tilde S^{\mu\nu}=0$, for this test. A
summation by parts numerical algorithm then gives rise to
the semi-discrete conservation law
\begin{equation}
 \partial_t \sum_{I,J,K}  \left( g^{t\beta} \partial_\beta \bar g^{\mu\nu} \right) = 0.
\label{eq:gaugewave.abigel.cons}
\end{equation}
While this is a non-generic result (most space-times would give a non-zero
source term),  building this conservation law into the principal part of the
system has proved effective not only in this particular case but in the other
Apples with Apples tests considered in this paper, as well as in further
proposed tests~\cite{Babiuc:2004pi,Babiuc-etal-2005,Babiuc:2006wk}.

\begin{figure}[hbtp]
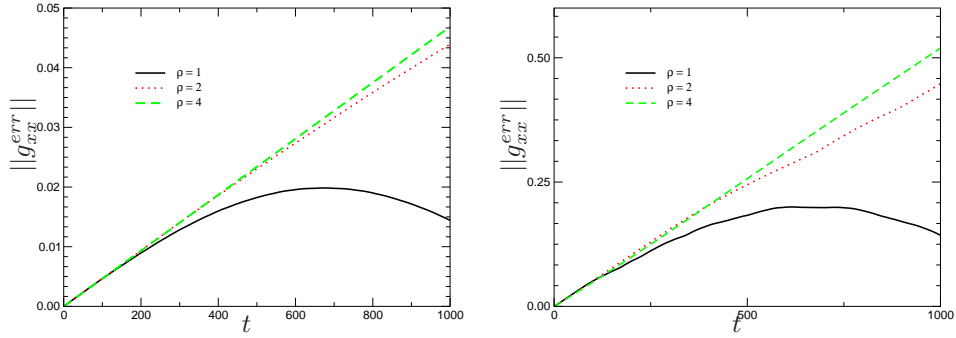

\begin{center}
  \psfrag{xlabel}{$t$}
  \psfrag{ylabel}{$||g^{err}_{xx}||$}
  \includegraphics*[width=14pc]{mex1resfig08}\hspace{5mm}
  \includegraphics*[width=14pc]{mex1resfig09}
\end{center}
  \caption{Convergence results for the 1D gauge wave simulation with the
Abigel\_harm code, for amplitudes of $A=0.01$ (left) and $A=0.1$ (right). The
graphs show the $L_\infty$ norm of the error in $g_{xx}$, defined as
$g^{err}_{xx} = g^{num}_{xx} - g^{ana}_{xx}$ as a function of time, and rescaled
by a factor of $1/\rho^2$. As seen from the graphs, the lower amplitude runs
give no new information.
}
\label{fig:Abigel.gw1D.gxx_e}
\vspace{3ex}
\end{figure}

\begin{figure}[hbtp]
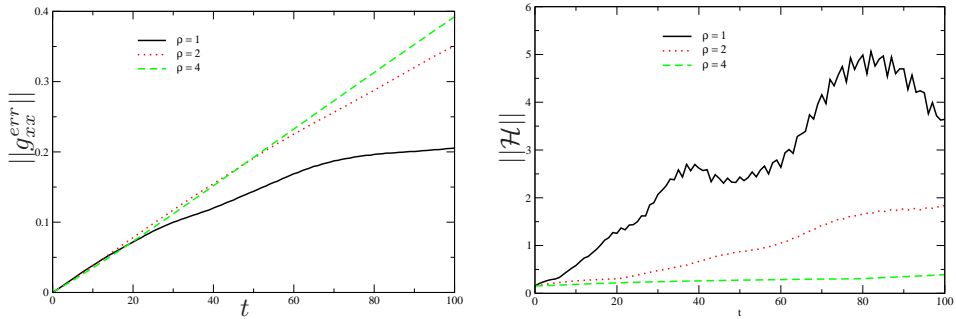

\begin{center}
  \psfrag{xlabel}{$t$}
  \psfrag{ylabel}{$||g^{err}_{xx}||$}
  \includegraphics*[width=14pc]{mex1resfig10}\hspace{5mm}
  \psfrag{xlabel}{$t$}
  \psfrag{ylabel}{$||\cal{H}||$}
  \includegraphics*[width=14pc]{mex1resfig11}
\end{center}
  \caption{Convergence results for the 2D gauge wave simulation with the
Abigel\_harm code, for amplitude $A=0.5$.  The left graph shows the $L_\infty$
norm of the error in $g_{xx}$, rescaled by a factor of $1/\rho^2$, as a function
of time; while the right graph shows the same rescaled error norm for the
violation of the Hamiltonian constraint ${\cal H}$. For the Abigel\_harm code,
the vanishing of the Hamiltonian constraint is an algebraic identity, making
${\cal H}$ of order roundoff. As a result, the constraint violation is
super convergent. The lower amplitude runs revealed no new features.
}
\label{fig:Abigel.gw2D.err}
\end{figure}

As shown in Figs. \ref{fig:Abigel.gw1D.gxx_e} and
\ref{fig:Abigel.gw2D.err}, the flux-conservative code does not develop
exponential error modes
when running
with the original ICN integrator (see \cite{Babiuc-etal-2005} for similar
results with
RK4.)  The main source of error is phase error
which converges to zero as the grid is refined. In order to further illustrate this point, Figs.
\ref{fig:Abigel.gw1D.gxx_e} and \ref{fig:Abigel.gw2D.err} give test
results for both the 1D and 2D versions with amplitudes of $A=0.01, \; 0.1, \;
0.5$.

%

%%%%%%%%%%%%%%%%%%%%%%%%%%%%%%%%%%%%%%%%%%%%%%%%%%
\subsubsection{Results for the HarmNaive System}
%%%%%%%%%%%%%%%%%%%%%%%%%%%%%%%%%%%%%%%%%%%%%%%%%%

This naive harmonic system, although weakly hyperbolic, behaves identical to the
symmetric hyperbolic Abigel\_harm code for this testbed. This can be understood
given that the RHS for the mixed space-time components of the evolution system
vanish, i.e.
\begin{equation}
\partial_t \bar g^{it} = - \partial_j \bar g^{ij} = 0,
\end{equation}
which implies that the time-time component of the RHS also vanishes, i.e.,
\begin{equation}
\partial_t \bar g^{tt} = - \partial_j \bar g^{tj} = 0.
\end{equation}
The test-results confirm this.

As expected, tests for the ADM-system also behave identically, since the
naive harmonic system can be understood as a formulation of the ADM-system
in the harmonic gauge. We therefore skip a separate discussion of the
ADM-system.

%%%%%%%%%%%%%%%%%%%%%%%%%%%%%%%%%%%%%%%%%%%%%%%%%%
\subsection{Results for the KrancFN and KrancNOR Systems}
%%%%%%%%%%%%%%%%%%%%%%%%%%%%%%%%%%%%%%%%%%%%%%%%%%

\begin{figure}[hbtp]
\begin{center}
  \psfrag{xlabel}{$x$}
  \psfrag{ylabel}{$g_{xx}$}
  \includegraphics*[width=20pc]{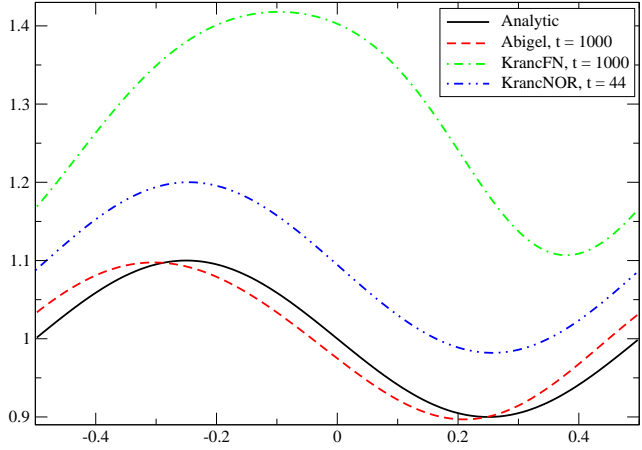}
\end{center}
  \caption{Comparison snapshots of $g_{xx}(x)$ for the 1D gauge wave with
amplitude $A=0.1$ at the end of a run with $\rho =4$ resolution. For the Abigel
and KrancFN codes, the run lasts the full 1000 crossing times. The KrancNOR code
crashes at $t=44$. 
}
\label{fig:GWKrancsurface}
\end{figure}

Besides the harmonic codes, KrancFN was the only other code that was able to run
for 1000 crossing times for an amplitude $A=.1$. At the end of the run,
Fig.~\ref{fig:GWKrancsurface} shows that long wavelength growth due to
the $e^{\lambda t} H$ instability  of the wave amplitude has become appreciable.

The KrancNOR code picks up the $e^{\lambda t} H$ instability at a faster rate
and, although it shows clear 2nd order convergent at early times, it crashes at
$t\approx 44$. The snapshot in Fig.~\ref{fig:GWKrancsurface} shows
that the error at the end of the run is almost exactly in the $e^{\lambda t} H$
mode. 

%

%%%%%%%%%%%%%%%%%%%%%%%%%%%%%%%%%%%%%%%%%%%%%%%%%%
\section{Shifted gauge wave testbed}
\label{sec:shiftedtest}
%%%%%%%%%%%%%%%%%%%%%%%%%%%%%%%%%%%%%%%%%%%%%%%%%%

In the shifted gauge wave test (\ref{eq:sgw}) we have identified two types of
instability~\cite{Babiuc-etal-2005}. One, which is analogous to the instability
of the gauge wave, arises from the $\lambda$-parameter family of vacuum metrics
\begin{equation}
  ds_\lambda^2=e^{\lambda t}(- dt^2 +dx^2)+dy^2+dz^2
           +H k_\alpha k_\beta dx^\alpha dx^\beta,
\label{eq:lsgw}
\end{equation}
which reduces to the shifted gauge wave for $\lambda=0$.
The other is an instability peculiar to harmonic
(or generalized harmonic) evolution codes, where the Einstein
equations are satisfied only indirectly through the
harmonic conditions. The metric
\begin{equation}
  d\hat s_\lambda^2=- dt^2 +dx^2+dy^2+dz^2 +
        \bigg(H-1 +e^{\lambda \hat t}\bigg )
            k_\alpha k_\beta dx^\alpha dx^\beta ,
\label{eq:instab}
\end{equation}
where
\begin{equation}
\hat t= t - \frac {Ad}{4\pi}\cos \left( \frac{2 \pi (x - t)}{d} \right) ,
\end{equation}
satisfies the reduced harmonic evolution equations
(\ref{eq:gaugewave.abigel.2nd}). The simulation of the shifted gauge wave by
any evolution code based upon a standard reduction of Einstein's
equations to harmonic form can be expected to excite this instability. 

\begin{figure}[hbtp]
\begin{center}
  \psfrag{xlabel}{$x$}
  \psfrag{ylabel}{$g_{xx}$}
  \includegraphics*[width=20pc]{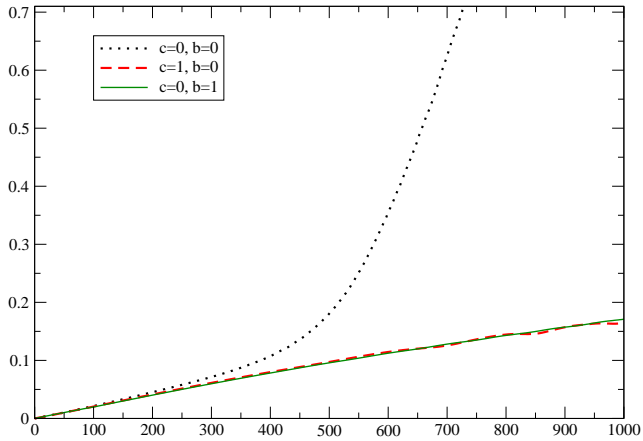}
\end{center}
  \caption{Plots of the $L_\infty$ error ${\cal E}(t)$ in $g_{xx}$ obtained with
the Abigel code for the 1D shifted gauge wave test with amplitude $A=.5$ and
resolution $\rho = 4$.  Results are compared for the  constraint adjustment
(\ref{eq:mmmmbeladj}) with b = 1, the constraint  adjustment (\ref{eq:beladj})
with c = 1 and the bare algorithm.  The two adjustments show very similar error
and both give excellent  suppression of the unstable mode excited by the bare
algorithm.
}
\label{fig:AbigelAGW1Dshcomp}
\end{figure}

The test was developed in conjunction with the Abigel\_harm
code~\cite{Babiuc-etal-2005}. For 1D runs with the $\rho=4$ resolution, it was
found that the evolution equation (\ref{eq:gaugewave.abigel.FC}) excited the
instability (\ref{eq:instab}) on a timescale $t\approx 500$. Further
investigation showed that this instability could be suppressed by adjusting
(\ref{eq:gaugewave.abigel.FC}) according to
\begin{equation}
   \tilde S^{\mu\nu}\rightarrow \tilde S^{\mu\nu}-A^{\mu\nu},
\end{equation}
where $A^{\mu\nu}=0$ when the harmonic constraints
\begin{equation}
  C^\mu:=-\frac{1}{\sqrt{-g}}(\partial_\nu{\bar g}^{\mu\nu}-\tilde H^\mu) =0
\end{equation}
are satisfied. Particularly effective were the constraint adjustments
\begin{equation}
      A^{\mu\nu}= \frac{b {\cal C}^\alpha \nabla_\alpha t}
      {e_{\rho\sigma}{\cal C}^\rho {\cal C}^\sigma}
         {\cal C}^{\mu} {\cal C}^{\nu}, \quad  b>0 ,
\label{eq:mmmmbeladj}
\end{equation}
where $e_{\rho\sigma}$ is the natural metric of signature $(++++)$ associated
with the Cauchy slicing, and
\begin{equation}
      A^{\mu\nu}= -\frac {c}{\sqrt{-g}} {\cal C}^\alpha 
              \partial_\alpha (\sqrt{-g}g^{\mu\nu}) , \quad  c>0 . 
\label{eq:beladj}
\end{equation} This is exhibited in Fig.~\ref{fig:AbigelAGW1Dshcomp}, which 
shows for a run with amplitude $A=0.5$ that these constraint adjustments 
suppress instabilities for the entire 1000 crossing time duration of the test.

\begin{figure}[hbtp]
\begin{center}
  \psfrag{ylabel}{$||{\cal H}||$}
  \psfrag{ylabel2}{$\Delta||K||$}
  \psfrag{xlabel}{$t$}
  \includegraphics*[width=14pc]{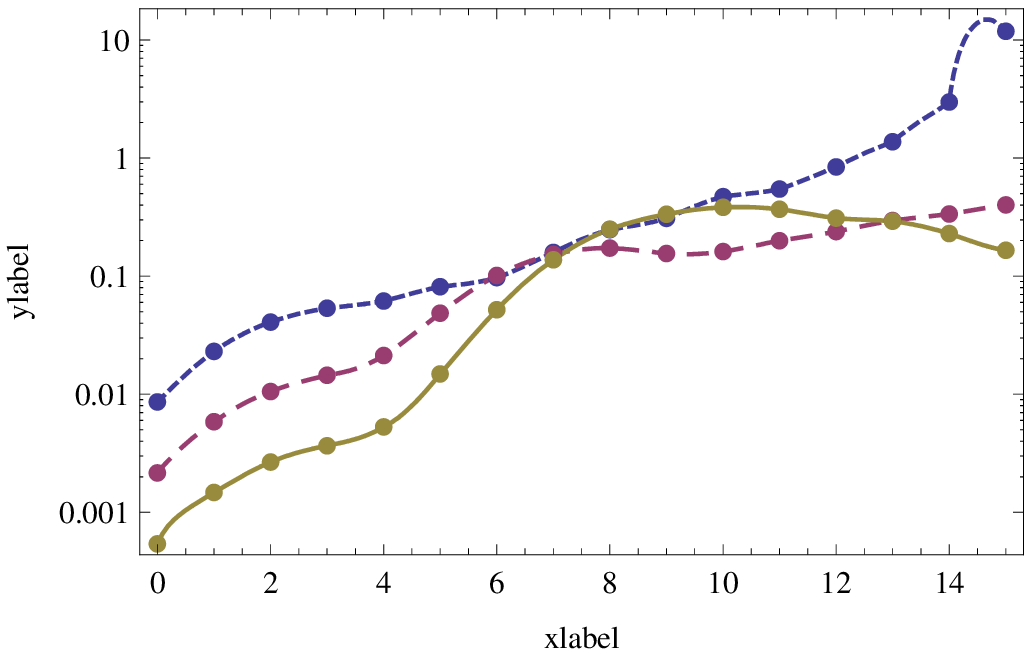}\hspace{5mm}
  \includegraphics*[width=14pc]{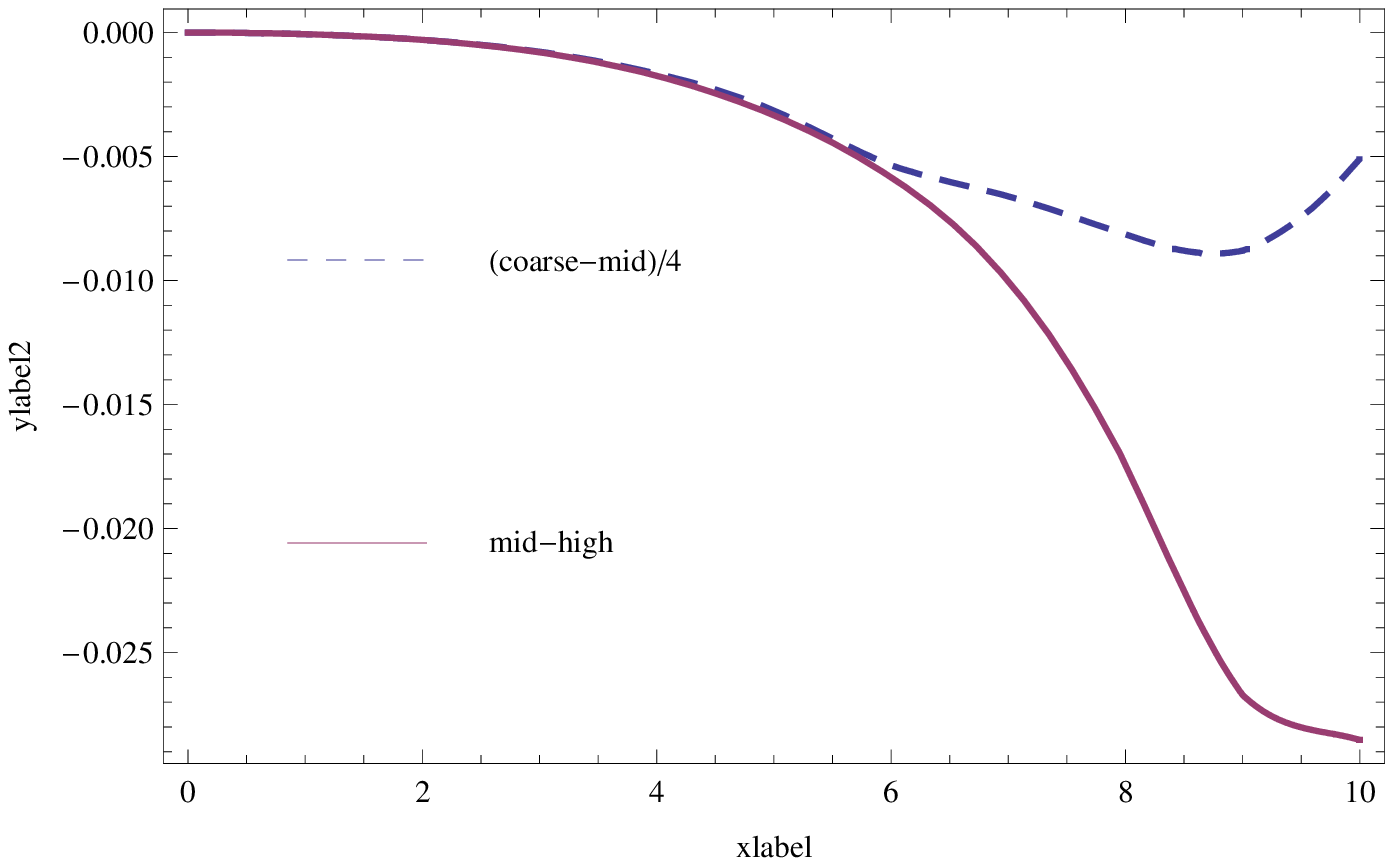}
\end{center}
 \caption{Performance of Kranc\_BSSN for the shifted gauge wave
with amplitude $A=0.1$ and a dissipation value of $\sigma=0.001$.
Left panel: The $L_2$-norm of the Hamiltonian constraint plotted
vs. time for resolutions $\rho=1,2,4$ (short-dashed, long-dashed, full line). 
Right panel: Convergence test for the $L_2$-norm of $trK$
for resolutions $\rho=1,2,4$. Second order convergence is lost 
after a few crossing times.
}
\label{fig:BSSN_ShiftedGW}
\end{figure}

Results for the shifted gauge wave tests are also available from the CVS
repository for BSSN codes. In this case, as in the standard gauge wave test,
results are not satisfactory, and suggest further analysis, which is beyond the
scope of this paper. Results obtained with the Kranc\_BSSN code and a very small
value of the dissipation parameter ($\sigma=0.001$, see
Eq.~(\ref{eq:KOdissipation})) for  the medium amplitude $A=0.1$ are shown in
Fig.~\ref{fig:BSSN_ShiftedGW}. While the code shows second order convergence for
several crossing times, rather quickly an instability develops that eventually
crashes the code. As expected, the instability develops slower for the lower
amplitude $A=0.01$, and faster for $A=0.5$, where the code crashes within
roughly one crossing time. Similar results are also available in the  CVS
repository for the CCATIE code. 

Results for the shifted gauge wave test have also been obtained~\cite{Boyle07}
using the Caltech-Cornell group's spectral version of a code based upon the
Kidder-Scheel-Teukolsky formulation of the Einstein equations~\cite{Kidder01a}.
For the 1D test with $A=.5$, they encountered nonlinear instabilities associated
with aliasing after a few crossing times. There are standard filtering
techniques to deal with such aliasing problems. By filtering the top 1/3
spectral coefficients, they found that the evolutions could be extended as far
as $t=60$, but further improvements by filtering did not seem possible. The
group has not yet reported results for their current spectral code 
which is based upon a generalized harmonic formulation.

%%%%%%%%%%%%%%%%%%%%%%%%%%%%%%%%%%%%%%%%%%%%%%%%%%
\section{Gowdy wave test}\label{sec:gowdywavetest}
%%%%%%%%%%%%%%%%%%%%%%%%%%%%%%%%%%%%%%%%%%%%%%%%%%

The previous tests involve spacetimes with small curvature. The Gowdy wave
test is based upon a strongly curved exact solution for an expanding
vacuum universe containing a plane polarized gravitational wave
propagating around a 3-torus
$T^3$~\cite{Gowdy71}. See~\cite{Ringstrom03}
for a recent review. The metric has the form
\begin{equation}
  {ds}^2 = t^{-1/2} e^{\lambda/2} (-{dt}^2 + {dz}^2) 
     + t(e^P {dx}^2 + e^{-P}{dy}^2),
  \label{1metric-gowdy}
\end{equation}
where $P(t,z)$ and $\lambda(t,z)$ depend periodically on $z$ and the time
coordinate $t$ increases as the universe expands, with a cosmological type
singularity at $t=0$.  The detailed tests specifications given in
\ref{sec:rgowdy} were designed so that neither very large nor very small numbers
enter in the initial data.

In the expanding direction, the qualitative behavior of the solution is
characterized by $P$ slowly decaying  to zero while $\lambda$ grows
linearly, with both $P$ and $\lambda$ exhibiting gravitational wave
oscillations. The linear growth of $\lambda$ leads to exponential
growth of $g_{zz}$, so that code accuracy is tested in a  harsh situation.
This makes evolution with a 3D code difficult  compared with the direct 1D
evolution of $P$ used in numerical studies of the approach to the
cosmological singularity~\cite{Berger02}

\begin{figure}[hbtp]
\begin{center}
  \psfrag{xlabel}{$t$}
  \psfrag{ylabel}{$K$}
  \includegraphics*[width=20pc]{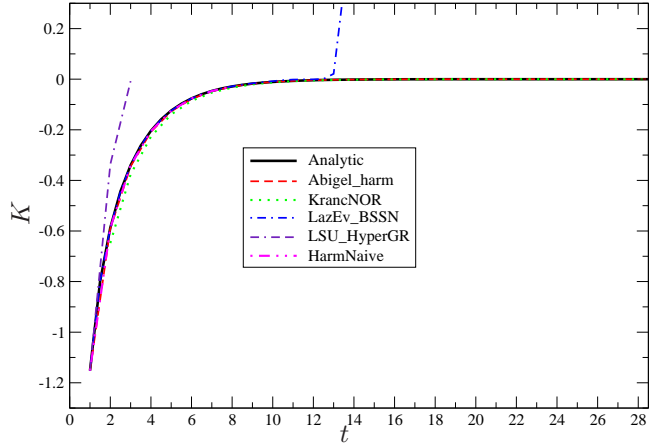}
\end{center}
  \caption{Comparison plots  of the trace of the extrinsic curvature $K$ for
the polarized Gowdy wave evolved in the expanding direction with the
$\rho=4$ resolution. Analytically $K$ is spatially homogeneous; the plots
show its maximum value over the numerical grid.
}
\label{fig:PGowdyExptrKmax}
\end{figure}

\begin{figure}[hbtp]
\begin{center}
  \psfrag{xlabel}{$z$}
  \psfrag{ylabel}{${\cal E}$}
  \includegraphics*[width=20pc]{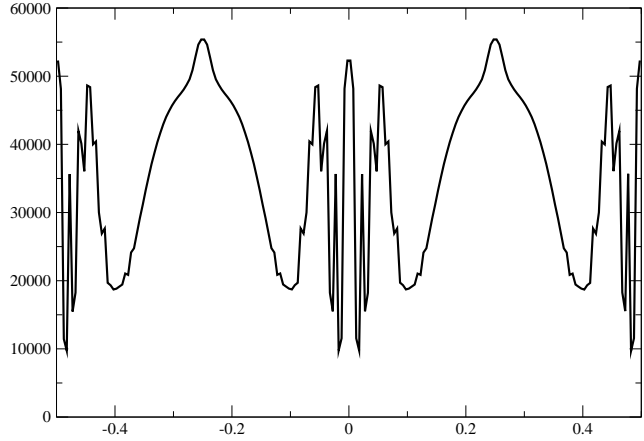}
\end{center}
  \caption{Plot of the error ${\cal E}(z)$ in $g_{zz}$ for
the polarized Gowdy Wave evolved in the expanding direction with  $\rho=4$
resolution with the \em$2^{nd}$\em order accurate LazEv\_BSSN code. The
error, plotted at $t=13$ just before the code crashes, shows a large short
wavelength component which can be controlled by dissipation.
} 
\label{fig:PGowdyExpgzze_s13}
\vspace{3ex}
\end{figure}

The performance of the various codes in the expanding direction is illustrated
by the output for the trace of the extrinsic curvature $K$ shown in
Fig.~\ref{fig:PGowdyExptrKmax}. Although not apparent from the figure, the
HarmNaive code crashes abruptly at $t=8$, as might be expected of a weakly
hyperbolic system in the nonlinear regime. Even though the analytic value of $K$
is negative and asymptotes to zero with the expansion, short wavelength error in
the LS\_HyperGR and LazEv\_BSSN codes triggers an instability leading to a
collapsing mode with $K >0$.  This is illustrated for the LazEv\_BSSN run  in
the snapshot of Fig.~\ref{fig:PGowdyExpgzze_s13}, which shows the  error in
$g_{zz}(t,x)$ at $t=13$  just before the run crashes. The superposition of short
wavelength error with the long wavelength truncation error from the signal is
evident.

\begin{figure}[hbtp]
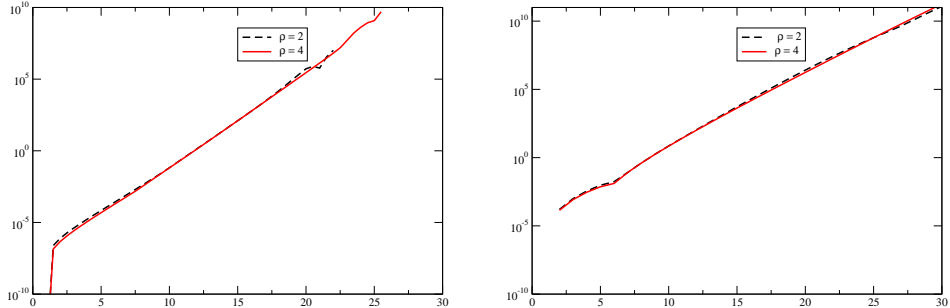

\begin{center}
  \psfrag{xlabel}{$t$}
  \psfrag{ylabel}{$||{\cal E}||$}
  \includegraphics*[width=14pc]{mex1resfig18} \hspace{5mm}
  \includegraphics*[width=14pc]{mex1resfig19}
\end{center}
  \caption{Convergence results for the $L_\infty$ norm of the error $||{\cal
E}(t)||$ in $g_{zz}$ (logarithmic scale) for the polarized Gowdy wave evolved in
the expanding direction. On the left, the results for the $\rho=2$ resolution
have been rescaled by 1/16 for the \em$4^{th}$\em order accurate LazEv\_BSSN
code with \em  dissipation\em. The results indicate stability and convergence
but do not give long term accuracy. On the right, the error for the $\rho=2$
resolution has been rescaled by 1/4  for the \em$2^{nd}$\em order system
Abigel\_Harm code, again showing stability and convergence. Both codes exhibit
roughly the same long term rate of error growth expected from the exponential
growth of $g_{zz}$.
} 
\label{fig:PGowdyExpgzze}
\end{figure}

Further experiments with the LazEv\_BSSN code showed that this short
wavelength instability could be controlled by numerical dissipation and that
the accuracy could be further improved by using fourth order finite
difference approximations. For the expanding Gowdy test, this is illustrated
in the plots of the rescaled error in the left portion of
Fig.~\ref{fig:PGowdyExpgzze} which indicate fourth order
convergence. However, the error still exhibits poor long term accuracy. In
the right portion of Fig.~\ref{fig:PGowdyExpgzze}, we also display the error
in the second order accurate Abigel\_Harm code. Both the second order and
fourth order codes have approximately the same long term rate of growth due
to the underlying exponential growth in $g_{zz}$.

The Gowdy test is run in {\em both} future and past time directions because
analytical studies \cite{Frauendiener:2004bj} and numerical experiments
\cite{Husa:2005ns,Husa02b} indicate that the sign of the extrinsic
curvature may have important consequences for constraint violation. The
subsidiary system governing constraint propagation can lead to unstable
departure from the constraint hypersurface. As an example, in a hyperboloidal
slicing of Minkowski space with unit lapse and zero shift, the electromagnetic
constraint $C=\nabla_a E^a$ satisfies $C(t)=C(0)e^{Kt}$ when the standard
Maxwell evolution equations are satisfied. Thus numerical error can be expected
to lead to exponential growth of the constraint for a hyperboloidal foliation
with $K>0$. The situation is more complicated in the nonlinear gravitational
case but similar instabilities of the system of equations governing the
constraints are associated with the extrinsic curvature
\cite{Frauendiener:2004bj}. A negative value of $K$ (the expanding case) tends
to damp constraint violation whereas a positive value (the collapsing case) can
trigger constraint violating instabilities.

\begin{figure}[hbtp]
\begin{center}
  \psfrag{xlabel}{$t$}
  \psfrag{ylabel}{$||\cal{H}||$}
  \includegraphics*[width=20pc]{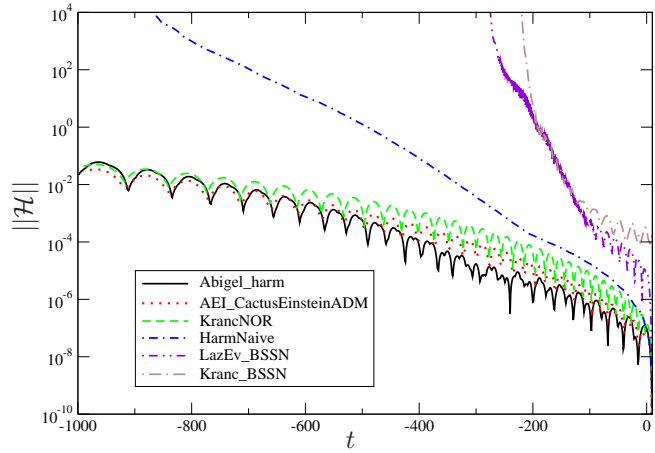}
\end{center}
  \caption{Comparison plot of the $L_\infty$
           norm of the Hamiltonian constraint vs harmonic time $t$ for the
           polarized Gowdy Wave evolved in the
           collapsing direction with the $\rho=4$ resolution.} 
\label{fig:PGowdyColhaminfcomp}
\end{figure}

\begin{figure}[hbtp]
\begin{center}
  \psfrag{xlabel}{$t$}
  \psfrag{ylabel}{$||\cal{H}||$}
  \includegraphics*[width=20pc]{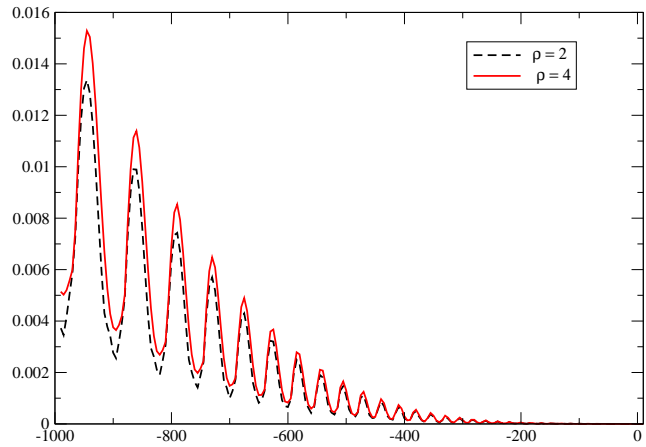}
\end{center}
  \caption{Convergence results for the $L_\infty$ norm of the Hamiltonian
constraint $\|{\cal H}(t)\|$ for the polarized Gowdy Wave evolved in the
collapsing direction by the \em$4^{th}$\em order system   LazEv\_BSSN code with
\em dissipation\em.  After rescaling the results for the $\rho=2$ by 1/16, they
closely match those for the $\rho=4$ resolution. The figure shows stability and
convergence of the Hamiltonian constraint up to 1000 crossing times and
demonstrates good performance of the LazEv\_BSSN code if  dissipation is added.
} 
\label{fig:LazEvDGowdyChammax}
\end{figure}

In the collapsing direction, we perform the runs with a harmonic time slicing to
prolong the approach to the singularity, as previously done by Garfinkle
\cite{Garfinkle02}.  Results for the Hamiltonian constraint for the various
codes are shown in Fig.~\ref{fig:PGowdyColhaminfcomp} for the collapsing case.
All the codes now show some growth in the Hamiltonian constraint, either of a
slow or runaway type. The slow growth, exhibited for example by the
Abigel\_harm, AEI\_CactusEinsteinADM and KrancNOR codes, can be attributed to
the analytic constraint instabilities discussed in \cite{Frauendiener:2004bj};
the Hamiltonian constraint violation remains small ($\approx 10^{-2}$) at the
end of the run. The runaway growth exhibited by the LazEv\_BSSN code can again
be controlled by numerical dissipation. This is demonstrated by the convergence
results shown in  Fig.~\ref{fig:LazEvDGowdyChammax} for the fourth order
dissipated version of the code.


The choice of specifications given in \ref{sec:rgowdy} provides a
Gowdy testbed capable of good discrimination between different formulations.
Results for both the expanding (Fig.~\ref{fig:PGowdyExptrKmax}) and
collapsing (Fig.~\ref{fig:PGowdyColhaminfcomp}) directions show 
a wide spread in the performance of the different codes. We observe, as in the
gauge wave test, that the BSSN-based codes have less satisfactory
performance.

%---------------------------------------------------------------------
%%%%%%%%%%%%%%%%%%%%%%%%%%%%%%%%%%%%%%%%%%%%%%%%%%
\section{Conclusions}\label{sec:discussion}
%%%%%%%%%%%%%%%%%%%%%%%%%%%%%%%%%%%%%%%%%%%%%%%%%%

This first round of tests, although modest in scope is a good start at
establishing the methods for code verification that have been deemed necessary
for any complicated computational discipline, such as numerical relativity, to
fulfill its scientific potential. As observed by Post and Votta~\cite{Post05} in
their study of the verification and validification of large scale computational
projects, ``the peer review process in computational science generally doesn't
provide as effective a filter as it does for experiment or theory. Many things
that a referee cannot detect could be wrong with a computational science
paper\ldots The few existing studies of error levels in scientific computer codes
indicate that the defect rate is about seven faults per 1000 lines of Fortran''.
Their observations are especially pertinent for numerical relativity where
validation by agreement with experiment is not available. 

Several problems have been encountered in the course of this project. One
problem was getting  prompt response from a broad set of groups with many other
pressures. The Apples with Apples workshops were very successful in this regard
and were absolutely essential in jump-starting and continuing the project. But
after the participants dispersed from the workshops, outside pressures  led to
predictable difficulties. Besides teaching and administrative duties, the
overriding scientific pressure in the field has been solving the two black hole
problem and supplying waveforms.  This raises a complicated juggling of
priorities between black hole simulations and code verification. In order for
code verification to be attractive, the tests have to be useful and the
investment in time has to be minimal. This adds emphasis on the need for
tests that are simple to carry out and simple to document the results.

Another level of complication in this project arises from the feedback between
test design and the analysis of test output. This has led us to
improvements in the tests and to their better understanding. In the robust
stability test the correct interpretation of results for weakly hyperbolic
algorithms required rethinking the proper choice of norm and refinement
procedure for judging stability. In the gauge wave tests, the desire for
computational efficiency in detecting nonlinear problems at an early time has
led us to the adoption of a higher amplitude $A=0.5$ for the test, as
opposed to the original specifications $A=0.01$ and $A=0.1$. 

The robust stability test is presented as a pass/fail test. For the linear wave
test the amplitude and phase errors in the output data for the wave profile
provide a good comparison of code performance. For the gauge wave and shifted
gauge wave tests, a prime challenge is the suppression of long wavelength
nonlinear instabilities in the analytic problem. For the Gowdy test, there were
unanticipated shortcomings in the output content that should lend valuable
experience in the design of future black holes tests. Useful benchmarks have
been established for the linear wave, gauge wave, and Gowdy wave tests, which
have revealed clear deficiencies in various codes. Such deficiencies raise a
clear alert that it is necessary to apply or recheck other verification
techniques, such as convergence tests.

These first round results provide a good basis for proposing new tests. Already,
they have prompted addition of the shifted version of the gauge wave test, in
which  a non-vanishing shift fills a gap in the four original tests for periodic
boundary conditions. This test has been useful in developing analytic and
numerical techniques for controlling
instabilities~\cite{Babiuc-etal-2005,Boyle07}. A second round of
boundary tests based upon
the periodic tests have been proposed. The specifications are given on the
Alliance website~\cite{applesweb}. Results of some of these boundary tests have
been reported elsewhere~\cite{Babiuc:2006wk,Rinne06}. The next stage is to
formulate tests involving black holes.

The code comparisons have proved useful for designing code improvements and for
stimulating the use of new numerical techniques. During the course of this work,
results of the shifted gauge wave test were key to recognizing the importance of
discrete energy and flux conservation for harmonic code
performance~\cite{Babiuc-etal-2005}. The need to carry out the tests with a wide
range of formulations has led to the development of symbolic code generation
\cite{Husa:2004ip}. Although the tests were designed for finite difference
codes, they have been adapted and applied to pseudo-spectral
codes~\cite{Boyle07}. Further independent studies based upon the tests have
played a major part in thesis research \cite{Alic05,Hinder2005}.

Establishment of the CVS data repository has been an important step in the
documentation of test results.  Instructions for accessing the data are given
at~\cite{applesweb}. The CVS directory structure has been significantly
streamlined and documented since the beginning of the project. However, the
difficulties in completing this analysis of the first round of tests has
emphasized the need of a uniform standard for data structures and output.
Rather than trying to anticipate a complete list of useful output quantities,
it seems more desirable to output the 3-metric and extrinsic curvature at
specified times. Then other output quantities can be constructed in post
processing. Ideally, this should be done in some standardized way using
automated routines and graphical interfaces. All of this would require
considerable infrastructure to provide hardware for data storage and software
for processing. This is one of the important matters that will be presented for
discussion at future Alliance meetings.

\ack

We are grateful to UNAM in Mexico City, to the University of Cordoba, and to
the Center for Computation \& Technology at Louisiana State University for
their hospitality in conducting AwA workshops. M.C.B, S.H., C.L., E.S., and
J.W. gratefully acknowledge hospitality of the Albert Einstein Institute, and
S.H. of the University of the Balearic Islands.  We benefited from
discussions with Miguel Alcubierre, Adel Benlagra, Carles
Bona, Mihaela Chirvasa, Helmut Friedrich, Ian Hawke,  Frank Herrmann, 
Carlos Palenzuela, Oscar
Reula, Krzysztof Roszkowski, Marcelo Salgado, Hisa-aki Shinkai, Manuel
Tiglio, and Tilman Vogel.  The work of  M.C.B and J.W. was supported by NSF
grant PH-0553597 to the University of Pittsburgh. Y.Z. was supported by NFS
grants PHY-0722315, PHY-0722703, PHY-0714388, and PHY-653303, and by the NASA
Center for Gravitational Wave Astronomy at the University of Texas at
Brownsville (NAG5-13396). D.A. has been supported
by the Spanish Ministero de Education y Ciencia, projects FPA-2007-60220
and HA2007-0042, S.H. has been supported by DAAD through a PPP-project with
the University of the Balearic Islands and by DFG through the
SFB/TR7. 
S. Husa is a VESF fellow of the European Gravitational Observatory
(EGO).

\appendix

%%%%%%%%%%%%%%%%%%%%%%%%%%%%%%%%%%%%%%%%%%%%%%%%%%
\section{Revised testbed specifications}
\label{sec:revspecs}
%%%%%%%%%%%%%%%%%%%%%%%%%%%%%%%%%%%%%%%%%%%%%%%%%%

We present here the updated specifications for the five standardized testbeds.
For each test we provide the 4-metric of the spacetime, except for the robust
stability test where only the initial Cauchy data is specified. This determines
the 3-metric $h_{\mu\nu}=g_{\mu\nu}+n_\mu n_\nu$ (where $n_\mu$ is the future
directed unit normal to the Cauchy hypersurface) and the extrinsic curvature
$K_{\mu\nu}$. We use the convention $K_{\mu\nu}= -h_\mu^\rho \nabla_\nu n_\rho$
for which the trace $K$ is negative for an expanding cosmology. In all cases,
the evolution takes place in a fixed rectilinear coordinate domain with periodic
boundary conditions, i.e. a 3-torus. The identified ``boundaries'' in the
3-torus picture are located a half step from the first and last grid points
along each axis. 

Even though we are concerned with 3-dimensional codes, for tests
with only one-dimensional features
in the $x$-direction it
is efficient to use the minimum number of grid points in the 
trivial $y$ and $z$ directions, i.e. to run the test in a long channel rather
than a cube. 
For standard second order finite differencing this
implies that we use 3 or 4 points in those directions. 
For all such 1D tests, the evolution domain is
\begin{equation}
 \quad x \in [-0.5, +0.5],  \quad y = 0, \quad z = 0,
\end{equation}
with grid
\begin{equation} 
    x = -0.5 + (n-\frac{1}{2}) dx, \quad n=1\ldots 50\rho,
  \quad dx=1/(50\rho), \quad \rho \in \mathbb{Z}.
  \label{eq:grid}
\end{equation}
(In the Gowdy wave test, the grid is aligned with the $z$-direction.)
The 2D tests have evolution domain
\begin{equation}
  \quad x \in [-0.5; +0.5]
    \quad y \in [-0.5; +0.5],\quad z=0
\end{equation}
with both the $x$ and $y$ grids satisfying (\ref{eq:grid}). The parameter $\rho$ allows
for grid refinement.  The coarsest $\rho=1$ grid is useful only for debugging.
Convergence tests should be made with $\rho=2$ and $\rho=4$, with benchmarks for norms,
constraints, etc. provided by $\rho=4$.

We have dropped the original requirement that the tests be run with an
iterative-Crank-Nicholson algorithm since Runge-Kutta time integrators have
since proved to be more effective and have been commonly adopted. For each test,
the size of the timestep $dt$ is given in terms of the grid size to lie within
the CFL limit for an explicit evolution algorithm. (For some codes this may be
inappropriate and some equivalent choice of time step should be made.). A final
time $T$, and intermediate times for data output, are specified for each test.
They are chosen to incorporate all useful features of the test without
prohibitive computational expense. Except for the robust stability test, it is
important to calculate the convergence rate of the numerical error. Additional
output variables might be essential to assess the performance of a particular
formulation.

%%%%%%%%%%%%%%%%%%%%%%%%%%%%%%%%%%%%%%%%%%%%%%%%%%
\subsection{Robust stability testbed}
\label{sec:rrobust}
%%%%%%%%%%%%%%%%%%%%%%%%%%%%%%%%%%%%%%%%%%%%%%%%%%

The 3-metric is initialized as $h_{ij}=\delta_{ij}+\epsilon_{ij}$, where
$\epsilon_{ij}$ are independent random numbers at each grid point. All other evolution
variables are initialized in the same way. The amplitude of the random noise is scaled
with the grid as
\begin{equation} 
    \label{eq:RobustStabilityAmplitude} 
         \epsilon \in (-10^{-10}/\rho^2,+10^{-10}/\rho^2). 
\end{equation}
The range of the random numbers ensures
that $\epsilon^2$ effects are below round-off accuracy so that the
evolution remains in the linear domain unless instabilities arise.

The timestep is specified to be $dt = dx/10 = 0.002 / \rho$.
The use of 4 distinct gridpoints in the $y$ and $z$ directions allows for
instabilities associated with the
checkerboard mode.

The test should be run until one is confident that dissipation effects do not
cloud the result. Without artificial dissipation, a run time of one crossing
time, using output at every time step, is usually sufficient. This corresponds
to $500 \rho$ time steps. The test is passed if the norm satisfies the
inequality (\ref{def:stability}) for all resolutions, for some fixed choice of
constants $A$ and $K$. Appropriate norms for both first and second order systems
are recommended in~\cite{Calabrese:2005ft} and are publicly available as Cactus 
thorns~\cite{AEIThorns-Dissipation}.

%%%%%%%%%%%%%%%%%%%%%%%%%%%%%%%%%%%%%%%%%%%%%%%%%%
\subsection{Linear wave testbed}
\label{sec:rlinwave}
%%%%%%%%%%%%%%%%%%%%%%%%%%%%%%%%%%%%%%%%%%%%%%%%%%

The initial 3-metric and extrinsic curvature $K_{ij}$ are given by a
transverse, trace-free perturbation with components
\begin{equation}
  \label{eq:linearwave4metric}
  ds^2= - dt^2 + dx^2 + (1+H) \, dy^2 + (1-H)  \, dz^2,
\end{equation}
where
\begin{equation}
  H =  A \sin \left( \frac{2 \pi (x - t)}{d} \right).
  \label{eq:H}
\end{equation}
This describes a linearized plane wave traveling in the $x$-direction.
The wavelength is set to $d=1$ to match the periodicity of the
evolution domain. The
metric has lapse
$\alpha=1$ and shift $\beta^i=0$.
The nontrivial components of extrinsic curvature are
\begin{equation}
  \label{eq:linearwave4extcurv}
   K_{yy} =   -\frac{1}{2} \partial_t H, \quad
   K_{zz} = \frac{1}{2} \partial_t H.
\end{equation}

In order to test 2-dimensional effects, the rotation
\begin{equation}
  \label{eq:flatgaugewave1to2d}
x = \frac{1}{\sqrt{2}}(x^\prime - y^\prime), \qquad
y = \frac{1}{\sqrt{2}}(x^\prime + y^\prime) \, .
\end{equation}
leads to a wave propagating along a diagonal.
The resulting metric is a function of 
\begin{equation}
\sin \left( \frac{2 \pi (x' - y' - t \sqrt{2})}{d'} \right), 
\quad \textrm{where} \quad d' = d \sqrt{2} \, .
\end{equation}
To obtain the required periodicity of the evolution domain, we set
$d=1$ in the 1D simulation and $d'=1$ in the diagonal simulation. The
test should be run in both axis-aligned and diagonal form. 

The test is performed with amplitude $A = 10^{-8}$, so that quadratic terms are
of the order of numerical round-off.  The time step is set to $dt = dx/4 = 0.005
/ \rho$ As in the gauge wave case, the 1D evolution is carried out for $T=1000$
crossing times, i.e.~$2\times10^5\rho$ time steps , with output every 10
crossing times.  The 2D diagonal runs are carried out for $T=100$, with output
every crossing time.  The output quantities are the $L_{\infty}$ and
$L_2$ norms, the maxima and minima, and profiles along the $x$-axis through the
center of the grid of $g_{yy}$, $g_{zz}$, 
Hamiltonian constraint; and the $L_{\infty}$ error
norm for $g_{zz}$ (measuring the difference from the exact solution).

%%%%%%%%%%%%%%%%%%%%%%%%%%%%%%%%%%%%%%%%%%%%%%%%%%
\subsection{Gauge wave testbed}
\label{sec:rgwave}
%%%%%%%%%%%%%%%%%%%%%%%%%%%%%%%%%%%%%%%%%%%%%%%%%%

The test is based upon the 4-metric 
\begin{equation}
  \label{eq:flatgaugewave4metric}
  ds^2=(1-H)(-dt^2 +dx^2)+dy^2+dz^2,
\label{eq:gwmet}
\end{equation}
with $H$ given by (\ref{eq:H}),
which is obtained from the Minkowski metric
$ds^2 =- d\hat t^2+d\hat x^2+d\hat y^2+d\hat z^2$ by the
transformation 
\begin{equation}
  \label{eq:GaugeWave1}
  \begin{array}[c]{r c l}
   \hat t&=& t - \frac {Ad}{4\pi}\cos \left( \frac{2 \pi (x - t)}{d} \right), 
      \\
   \hat x&=& x + \frac {Ad}{4\pi}\cos \left( \frac{2 \pi (x - t)}{d} \right), 
      \\
     \hat y&=& y,    \\ 
     \hat z&=& z .
  \end{array}
\end{equation}
This describes a sinusoidal gauge wave of amplitude $A$ propagating
along the $x$-axis. The extrinsic curvature is 
\begin{eqnarray}
  \label{eq:flatgaugewaveK}
  K_{xx} &=&\frac {\partial_t H}{2\sqrt{1-H}} =-\frac{\pi A}{d} \frac{ \cos \left( \frac{2 \pi (x - t)}{d}
  \right) }{ \sqrt{1 - A \sin \left( \frac{2 \pi (x - t)}{d} \right) }
  },\\
K_{ij} &=& 0 \qquad \textrm{ otherwise}.
\end{eqnarray}
As for the linear wave, the rotation (\ref{eq:flatgaugewave1to2d}) leads to wave
propagation along a diagonal with periodic boundary conditions.

The gauge wave is run with amplitude $A=.5$.  The time coordinate $t$ in the
metric (\ref{eq:flatgaugewave4metric}) is harmonic and the gauge speed is the
speed of light. The time step is set to $dt = dx/4 = 0.005 / \rho$. The
1D evolution is carried out for $T=1000$ crossing times, i.e.~$2\times10^5\rho$
time steps (or until the code crashes), with output every 10 crossing times. 
The 2D diagonal runs are carried out for $T=100$, with output every crossing
time. 

Output consists of the $L_{\infty}$ and $L_{2}$ norms, the maxima and minima,
and profiles along the $x$-axis through the center of the grid $(y=z=0)$ of
$g_{xx}$, $\alpha$, $tr(K)$ and the Hamiltonian constraint; and the $L_{2}$
error-norm for $g_{xx}$.

%%%%%%%%%%%%%%%%%%%%%%%%%%%%%%%%%%%%%%%%%%%%%%%%%%
\subsection{The shifted gauge wave test}
\label{sec:sgw}
%%%%%%%%%%%%%%%%%%%%%%%%%%%%%%%%%%%%%%%%%%%%%%%%%%

The shifted gauge wave is obtained from the Minkowski metric $ds^2 =-
d\hat t^2+d\hat x^2+d\hat y^2+d\hat z^2$ by the harmonic coordinate
transformation 
\begin{equation}
  \begin{array}[c]{r c l}
   \hat t&=& t - \frac {Ad}{4\pi}\cos \left( \frac{2 \pi (x - t)}{d} \right), 
      \\
   \hat x&=& x - \frac {Ad}{4\pi}\cos \left( \frac{2 \pi (x - t)}{d} \right), 
      \\
     \hat y&=& y,    \\ 
     \hat z&=& z
  \end{array}
\end{equation}
which leads to the Kerr-Schild metric
\begin{equation}
  ds^2=- dt^2 +dx^2+dy^2+dz^2 +H k_\alpha k_\beta dx^\alpha dx^\beta
\label{eq:sgw}
\end{equation}
where
\begin{equation}
      k_\alpha=-\partial_\alpha (t-x) 
\end{equation}
and $H$ is again given by (\ref{eq:H}). 
The extrinsic curvature is
\begin{eqnarray}
  \label{eq:flatsgaugewaveK}
  K_{xx} &=&\frac {\partial_t H}{2\sqrt{1+H}} ,\\
     K_{ij} &=& 0 \qquad \textrm{ otherwise}.
\end{eqnarray}
This metric describes a shifted gauge
wave of amplitude $A$ propagating along the $x$-axis.  The
coordinate transformation (\ref{eq:flatgaugewave1to2d}) rotates the propagation
direction to the diagonal. 

The shifted gauge wave test is run in a harmonic gauge with amplitude $A=0.5$ in
both 1D form and diagonal 2D form. As in the linear wave test, for
the required periodicity we set $d=1$ in the 1D simulations and $d'=1$ in the 2D
simulations.  We set the timestep $dt = dx/4 = 0.005 / \rho$. The 1D evolution
is carried out for $T=1000$ crossing times, i.e.~$2\times10^5\rho$ time steps
(or until the code crash). The 2D runs are carried out for $T=100$.

Output data consist of the profiles along the $x$-axis through the center of
the grid $(y=z=0)$ of $g_{tt}$, $g_{xt}$, and $g_{xx}$, the $L_2$ and
$L_\infty$ norms of the error and of the Hamiltonian constraint.

%%%%%%%%%%%%%%%%%%%%%%%%%%%%%%%%%%%%%%%%%%%%%%%%%%
\subsection{Polarized Gowdy wave testbed}
\label{sec:rgowdy}
%%%%%%%%%%%%%%%%%%%%%%%%%%%%%%%%%%%%%%%%%%%%%%%%%%

The polarized Gowdy metrics describe an
expanding, toroidal universe containing plane polarized gravitational
waves with metric
\begin{equation}
  {ds}^2 = t^{-1/2} e^{\lambda/2} (-{dt}^2 + {dz}^2) 
    + te^P {dx}^2 + e^{-P}{dy}^2,
  \label{metric-gowdy}
\end{equation}
where $\lambda$ and $P$ are
functions of $z$ and $t$ only and are periodic in $z$.
The universe expands as $t$ increases. The test is carried
out in both the collapsing and expanding situations. The metric is singular at
$t=0$.

The Einstein equations
reduce to a single evolution equation
\begin{equation}
    P_{,tt} + t^{-1} \, P_{,t} - P_{,zz} = 0 \label{p_evolution}
\end{equation}
and the constraint equations
\begin{equation}
  \lambda_{,t} = t\, (P_{,t}^2 + P_{,z}^2)\label{lambda-hamiltonian-constraint}
\end{equation}
and
\begin{equation}
  \lambda_{,z} = 2 \, t \, P_{,z} \, P_{,t} \label{lambda-momentum-constraint}.
\end{equation}

The test is based upon the particular solution to (\ref {p_evolution})
\begin{equation}
  P = J_0(2\pi t)\cos(2\pi z), \label{def_P}
\end{equation}
where $J_n$ are Bessel functions. The metric and extrinsic curvature are
\begin{equation}
  g_{xx}=te^P,\ g_{yy}=te^{-P},\ g_{zz}=t^{-1/2}e^{\lambda/2},
  \label{g_ij-polarizedgowdy}
\end{equation}
\begin{eqnarray}
  K_{xx}&=&- \frac{1}{2} t^{1/4}
  e^{-\lambda/4}e^P(1+t P_{,t}),
  \nonumber \\
  K_{yy}&=&- \frac{1}{2} t^{1/4}
  e^{-\lambda/4}e^{-P}(1 - t P_{,t}),
  \label{K_ij-polarizedgowdy} \\
  K_{zz}&=& \frac{1}{4} t^{-1/4}
  e^{\lambda/4}(t^{-1}-\lambda_{,t}),
  \nonumber
\end{eqnarray}
with
\begin{equation}
  trK = - \frac{1}{4}t^{1/4}e^{-\lambda/4}
           ( 3t^{-1} + \lambda_{,t}) .
\end{equation}
The shift vanishes and the lapse is
\begin{equation}
  \alpha=\sqrt{g_{zz}} = t^{-1/4}e^{\lambda/4}.
  \label{alpha-polarizedgowdy}
\end{equation}
For the choice (\ref{def_P}), the constraints
(\ref{lambda-hamiltonian-constraint},\ref{lambda-momentum-constraint}) 
yield
\begin{equation}
  \label{eq:lambda}
  \begin{array}[c]{r c l}
  \lambda &=& -2\pi t J_{0}(2\pi t) J_{1}(2\pi t) \cos^{2}(2\pi z)
   + 2\pi^{2}t^{2} \bigl[J_{0}^{2}(2\pi t) + J_{1}^{2}(2\pi t)\bigr]
   \\
   & & \mbox{} - {\frac{1}{2}}
   \bigl\{ (2\pi )^{2}\bigl[J_{0}^{2}(2\pi )
   +J_{1}^{2}(2\pi )\bigr]-
   2\pi J_{0}(2\pi ) J_{1}(2\pi )\bigr\}.
  \end{array}
\end{equation}
While $P$ slowly
decays to zero, $\lambda$ undergoes linear growth due to the
cosmological expansion, and both $P$ and $\lambda$ exhibit
gravitational wave oscillations. 

The velocity of light is constant in the coordinates chosen in
(\ref{metric-gowdy}) so that, with a fixed spatial discretization $dz$, the
Courant condition is consistent with a fixed timestep $dt$. This makes the
gauge (\ref{metric-gowdy}) convenient for evolving in the {\em expanding}
direction by choosing the initial data from the
exact solution at $t=1$, which yields data of order unity. 

In the backward in time evolution, we
choose a harmonic time slicing $\tau$ which only
asymptotically reaches the singularity.
Starting with the metric (\ref{metric-gowdy}), the
slicing is obtained by
a transformation $t = F(\tau)$, where the
harmonic condition $\Box \, \tau = 0$ implies
$F(\tau) = k e^{c \tau}$. In order to start the collapse slowly, the 
free constants $c$ and $k$ are chosen so that the new lapse
satisfies $\hat \alpha = 1$
at the initial time $t=t_0$.
This is accomplished by
picking $t_0$ for which $J_0(2 \pi t_0) = 0$ so that (\ref{eq:lambda})
implies $\hat \alpha$ is independent of $z$.
Using
$$
\tau_0 = \frac{1}{c} \ln \left( \frac{t_0}{k} \right), \quad
\lambda(k e^{c \tau_0}, z) = \lambda_0
$$
we obtain
\begin{equation}
\hat \alpha_0 = c\; t_0^{3/4}\; e^{\lambda_0 / 4}.
\end{equation}
Given our requirement $\hat \alpha_0 = 1$, and choosing 
$t_0 = \tau_0$, i.e.\ $F(\tau_0) = \tau_0$, we get
\begin{equation}
c = t_0^{-3/4}\; e^{-\lambda_0 / 4}, \quad
k = t_0 e^{-c t_0}.
\end{equation}
We choose a particular value of $t_0$ such that the initial slice
is far from the cosmological singularity, but not so far that we have
to deal with extremely large numbers by picking the $20$th
zero of the Bessel function $J_0(2 \pi t_0)$, which yields $t_0 \sim
9.8753205829098$, corresponding to
$$
c \sim 0.0021195119214617, \quad k \sim 9.6707698127638.
$$

The time step is set to $dt = dz/4 = 0.005 / \rho$ with run time $T=1000$ or
until code crash. Output consists of the $L_{\infty}$ and $L_2$-norms, the
maxima and minima, and profiles along the $z$-axis through the center of the
grid of $g_{zz}$, $\alpha$, $tr(K)$ and the Hamiltonian constraint. We output
norms every crossing time, and profiles either every 10 crossing times or once
per crossing time, depending on the behavior of the simulation.  We also output
the $L_{\infty}$ error norms of the difference from the exact solution for $g_{xx}$
and $g_{zz}$ for the expanding direction.

%%%%%%%%%%%%%%%%%%%%%%%%%%%%%%%%%%%%%%%%%%%%%%%%%%
\section{Code descriptions}\label{sec:appendix_codes}
%%%%%%%%%%%%%%%%%%%%%%%%%%%%%%%%%%%%%%%%%%%%%%%%%%

%%%%%%%%%%%%%%%%%%%%%%%%%%%%%%%%%%%%%%%%%%%%%%%%%%
\subsection{Standard ADM: Kranc\_FreeADM, and AEI\_CactusEinsteinADM codes}
%%%%%%%%%%%%%%%%%%%%%%%%%%%%%%%%%%%%%%%%%%%%%%%%%%

The formulation of the Einstein equation by Arnowitt, Deser and Misner (ADM)
\cite{Arnowitt62} provides a standard notion for ``evolving'' space-time as an
initial value problem in general relativity, which was initially presented in a
Hamiltonian context. What is referred to as a ``standard ADM'' system in the
numerical relativity community is a reformulation due to York \cite{York79},
which one obtains by 3+1--decomposition of the Einstein tensor  (as opposed to
3+1--decomposition of the Ricci tensor in the original ADM version), or
equivalently by adding appropriate constraint terms to the evolution equations.
As pointed out by Frittelli \cite{Frittelli97a},  York's ``standard ADM'' system
does in particular have nicer properties regarding the constraint propagation
system. This system is particularly simple, has a long history in numerical
relativity and exhibits some typical problems. We therefore use it as the
starting point for our numerical comparisons. The evolution equations are
\begin{eqnarray}
\partial_t \gamma_{ij}&=&
-2\alpha K_{ij}+\nabla_i\beta_j+\nabla_j\beta_i 
 \label{admevo1}
\\
\partial_t K_{ij} &=&
\alpha R^{(3)}_{ij}+\alpha K K_{ij}-2\alpha K_{ik}{K^k}_j
-\nabla_i\nabla_j \alpha
\nonumber \\ &&
+(\nabla_i \beta^k) K_{kj} +(\nabla_j \beta^k) K_{ki}
+\beta^k \nabla_k K_{ij}
,  \label{admevo2}
\end{eqnarray}
and the constraint equations are
\begin{eqnarray}
{\cal H}={\cal H}^{ADM}&:=& R^{(3)}+K^2-K_{ij}K^{ij}, \label{admCH} 
\\
{\cal M}_i={\cal M}_i^{ADM}&:=& \nabla_j {K^j}_i-\nabla_i K,  \label{admCM}
\end{eqnarray}
where $(\gamma_{ij}, K_{ij})$ are the induced three-metric and the extrinsic
curvature, $(\alpha, \beta_i)$ are the lapse function and the shift covector,
$\nabla_i$ is the 3-dimensional covariant derivative  and $R^{(3)}_{ij}$ is the
3-dimensional Ricci tensor associated with $\gamma_{ij}$.

We have tested two implementations of the standard ADM system, the
code AEI\_CactusEinsteinADM,
which is freely available via the website \cite{Cactusweb}, and Kranc\_FreeADM
which is based on the Cactus Toolkit \cite{Cactusweb} and Kranc software
\cite{Husa:2004ip}. AEI\_CactusEinsteinADM uses a hardcoded ICN time update
scheme (see e.g.\ \cite{Calabrese:2005ft}), whereas  Kranc\_FreeADM uses a
method of lines (MoL) approach based on the  CactusMoL thorn (in practice,
RK3, RK4 and ICN (see e.g.\ \cite{Calabrese:2005ft}) have also been used, as indicated).
In all of these codes, spatial partial derivatives are  reduced to partial
derivatives of the 3-metric, i.e., all expressions such as Christoffel symbols
are  expanded out. Due to the absence of first-order variables, no further
ambiguities arise. Centered second and  fourth order discretization is used (see
\ref{sec:appendix_FD}), and third order Kreiss-Oliger dissipation is optionally
applied to all variables (see \ref{sec:appendix_dissipation}).

The hyperbolicity of the ADM free evolution scheme has been analyzed and 
found to be weakly hyperbolic with the type of gauge conditions that we use
\cite{Calabrese:2005ft}.
Since many of our tests are essentially 1D tests, where ADM yields good
results, we have also analyzed the hyperbolicity of ADM in 1D.
For simplicity of presentation we restrict ourselves to the linearized case.
Assuming propagation in the x--direction we obtain the following evolution
equations. For the off-diagonal components,
$$
\partial_t \gamma_{yz} = 2 K_{yz}, \quad  
\partial_t K_{yz} = \partial_{xx} \gamma_{yz}/2, \quad 
\partial_t K_{xy} = 0, \quad 
\partial_t K_{xz} = 0.
$$
The evolution equations for $\gamma_{xy}$ and $\gamma_{xz}$ are analogous
to the evolution equation for $\gamma_{yz}$. 
The fact that the evolution equations for $K_{xy}$ and $K_{xz}$ are trivial 
renders the evolution system for the off-diagonal components weakly hyperbolic,
see e.g.\ \cite{Calabrese:2005ft}.
For the diagonal components,
\begin{eqnarray}
\partial_t \gamma_{ii} &=& 2 K_{ii} \qquad (i=x,y,z),\\
\partial_t K_{xx} &=& \partial_{xx}\alpha 
         + \frac{1}{2}\partial_{xx}(\gamma_{yy} + \gamma_{zz}), 
\\ 
\partial_t K_{jj} &=& \frac{1}{2} \partial_{xx} \gamma_{jj} \qquad (j=y,z).
\end{eqnarray}

Considering for simplicity the densitized lapse case,  $\alpha = \sqrt{\gamma}$,
the evolution equation for $K_{xx}$ becomes
$$
\partial_t K_{xx} = \frac{1}{2}\partial_{xx} \gamma_{xx}
         + \partial_{xx}(\gamma_{yy} + \gamma_{zz})
$$
and one finds that
the diagonal subsystem is only weakly hyperbolic. However, within the 
subclasses of gauge wave ($\gamma_{yy} = \gamma_{zz} = 0)$ or linear wave 
($\gamma_{xx} = 0$) data, the 1D ADM system corresponds to copies of
the 1D wave equation and is therefore well-posed.

%%%%%%%%%%%%%%%%%%%%%%%%%%%%%%%%%%%%%%%%%%%%%%%%%%
\subsection{Abigel\_harm}\label{codeAbigel}
%%%%%%%%%%%%%%%%%%%%%%%%%%%%%%%%%%%%%%%%%%%%%%%%%%

The Abigel code developed in Pittsburgh is based upon a symmetric hyperbolic
formulation of the Einstein equations using generalized harmonic coordinates
satisfying the curved space wave equation
\begin{equation}
   \Box x^\alpha= \frac{1}{\sqrt{-g}}\partial_\mu
        (\sqrt{-g}g^{\mu\nu}\partial_\nu x^\alpha)
   =\frac{1}{\sqrt{-g}}\tilde H^\alpha (x^\beta,g_{\rho\sigma}),
\label{eq:harmconds}
\end{equation}
where $\tilde H^\alpha$ are harmonic source terms.
The original version of the evolution equations was~\cite{Szilagyi02a}
\begin{equation}
       {\bar g}^{\alpha \beta} \partial_\alpha \partial_\beta
                    {\bar g}^{\nu\mu} = S^{\mu\nu}
\label{eq:gaugewave.abigel.2nd}
\end{equation}
where the left hand side is the principle part and the right hand side contains
nonlinear first-derivative terms. Here ${\bar g}^{\mu\nu} = \sqrt{-g}
g^{\mu\nu}$, with $g=\det(g_{\mu\nu})=\det({\bar g}^{\mu\nu})$. and the harmonic
constraints $\partial_\nu{\bar g}^{\mu\nu}=\tilde H^\mu$ are used in the
Einstein equations to eliminate second derivatives in the source terms
$S^{\mu\nu}$. For further details concerning the formulation and its
implementation see  \cite{Szilagyi02a}.

 The code with which the tests were performed was constructed by rewriting
(\ref{eq:gaugewave.abigel.2nd}) in the flux conservative form
\begin{equation}
   \partial_\alpha \left( g^{\alpha\beta} \partial_\beta {\bar g}^{\mu\nu} \right)
 =
     \tilde S^{\mu\nu}.
\label{eq:gaugewave.abigel.FC}
\end{equation}
and reducing it to the first order in time form
\begin{eqnarray}
  \partial_t {\bar g}^{\mu\nu} &=& -\frac{{\bar g}^{ti}}{{\bar g}^{tt}}\partial_i
{\bar g}^{\mu\nu}
            + \frac{\sqrt{-g}}{{\bar g}^{tt}}Q^{\mu\nu}
\label{eq:gammat}\\
    \partial_t Q^{\mu\nu}  &=&
    - \partial_i \bigg(
    g^{ij} \partial_j {\bar g}^{\mu\nu} +
    g^{it} \partial_t {\bar g}^{\mu\nu}
    \bigg)
        + \tilde S^{\mu\nu}
        \\ &=&
    - \partial_i \bigg[ \bigg(
    g^{ij}
-\frac{g^{ti} g^{tj}}{g^{tt}} \bigg) \partial_j
{\bar g}^{\mu\nu} \bigg]
            -  \partial_i \bigg( \frac{g^{it}}{g^{tt}}Q^{\mu\nu}\bigg)
        + \tilde S^{\mu\nu} \\ &=&
        - \partial_i \bigg( h^{ij} \partial_j {\bar g}^{\mu\nu} \bigg)
            -  \partial_i \bigg( \frac{g^{it}}{g^{tt}}Q^{\mu\nu}\bigg)
        + \tilde S^{\mu\nu}
\label{eq:Qt}
\end{eqnarray}
in terms of the evolution variables $({\bar g}^{\mu\nu},Q^{\mu\nu})$, where
\begin{equation}
      Q^{\mu\nu}= g^{t\alpha}\partial_\alpha {\bar g}^{\mu\nu}
\end{equation}
and
$h^{ij} = g^{ij} - g^{it} g^{jt} / g^{tt}$ is the spatial 3-metric.
Centered derivatives are used to finite difference (\ref{eq:gammat})
and the source terms $\tilde S^{\mu\nu}$ in (\ref{eq:Qt}). The remaining part
of Eq.~(\ref{eq:Qt}) is finite-differenced as follows:
\begin{eqnarray}
\fl
g^{\alpha\beta}_{[I+1/2,J,K]} &=& \frac{A_{+x}
{\bar g}^{\alpha\beta}_{[I,J,K]}}{\sqrt{- A_{+x} g_{[I,J,K]}}}
+ {\cal O}(\Delta^2)
\\ \fl
h^{ij}_{[I+1/2,J,K]} &=&     g^{ij}_{[I+1/2,J,K]}
-\frac{g^{ti}_{[I+1/2,J,K]} \; g^{tj}_{[I+1/2,J,K]}}{g^{tt}_{[I+1/2,J,K]}}
\\ \fl
 \partial_x  \bigg( h^{xx} \partial_x
{\bar g}^{\mu\nu} \bigg)_{[I,J,K]} &=&
 D_{-x} \bigg(
    h^{xx}_{[I+1/2,J,K]} \; D_{+x}{\bar g}^{\mu\nu}_{[I,J,K]} \bigg)
+ {\cal O}(\Delta^2)
\\ \fl
 \partial_x  \bigg( h^{xy} \partial_y
{\bar g}^{\mu\nu} \bigg)_{[I,J,K]} &=&
 D_{-x} \bigg(
    h^{xx}_{[I+1/2,J,K]} \; A_{+x} D_{0y}{\bar g}^{\mu\nu}_{[I,J,K]} \bigg)
+ {\cal O}(\Delta^2)
\\ \fl
  \partial_x \bigg( \frac{g^{xt}}{g^{tt}}Q^{\mu\nu}\bigg)_{[I,J,K]}
  &=& D_{-x} \bigg( \frac{g^{xt}_{[I+1/2,J,K]}}{g^{tt}_{[I+1/2,J,K]}} \;
  A_{+x} Q^{\mu\nu}_{[I,J,K]} \bigg) + {\cal O}(\Delta^2)
\end{eqnarray}
where the averaging operator $A_{+x}$ is defined in~\ref{sec:appendix_FD}.
The code is evolved as a first differential order in time and
second order in space system with a 2-step iterated Crank-Nicholson algorithm
or 4th order Runge-Kutta integrator.

%%%%%%%%%%%%%%%%%%%%%%%%%%%%%%%%%%%%%%%%%%%%%%%%%%
\subsection{HarmNaive}\label{codeHarmNaive}
%%%%%%%%%%%%%%%%%%%%%%%%%%%%%%%%%%%%%%%%%%%%%%%%%%

The HarmNaive code is based upon harmonic coordinates but differs from the
Abigel\_harm code because the evolution system consists of only the 6 wave equations
 (\ref{eq:gaugewave.abigel.FC}) for the spatial components ${\bar g}^{ij}$.
The time components are propagated by the harmonic conditions (\ref{eq:harmconds}), i.e.
\begin{equation}
        \partial_{t}{\bar g}^{\alpha t} + \partial_{i}{\bar g}^{\alpha i}
              =\hat H^{\alpha}.
\label{eq:hupdate}
\end{equation}
The coupling between ${\bar g}^{ij}$ and ${\bar g}^{\alpha t}$ makes the system
only weakly hyperbolic.

The evolution equations for ${\bar g}^{ij}$ and $Q^{ij}$ are finite differenced
as in the Abigel\_harm code. The evolution equation (\ref{eq:hupdate}) for
${\bar g}^{\alpha t}$ is approximated by central differences.
The update scheme is a 2-step iterative Crank-Nicholson algorithm.

%%%%%%%%%%%%%%%%%%%%%%%%%%%%%%%%%%%%%%%%%%%%%%%%%%
\subsection{KrancNOR code}
%%%%%%%%%%%%%%%%%%%%%%%%%%%%%%%%%%%%%%%%%%%%%%%%%%

%%%%%%%%%%%%%%%%%%%%%%%%%%%%%%%%%%%%%%%%%%%%%%%%%%
\subsubsection{Continuum formulation:}
%%%%%%%%%%%%%%%%%%%%%%%%%%%%%%%%%%%%%%%%%%%%%%%%%%

\newcommand{\fDef}
{\gamma^{kl}(\gamma_{ik,l} - \frac{1}{2} \rho \gamma_{kl,i})}

\newcommand{\KadditionA}[0] { \frac{a}{2} G_{(i,j)} }

\newcommand{\KadditionB}[0]
{
(c {\mathcal H} + a' G_{k,l}  \gamma^{kl}) \gamma_{ij}
}

Nagy, Ortiz and Reula suggested \cite{Nagy:2004td} modifications to
the ADM system such that it can be made strongly hyperbolic whilst
remaining in second order form.  The system we use includes slight
adjustments of \cite{Gundlach:2004jp}. Additionally, we use an evolved
lapse.

The variable  $f_i$ is defined as
\begin{eqnarray}
f_i = \fDef
\label{fEvolve}
\end{eqnarray}
with parameter $\rho$.  This introduces the new constraint $G_i$ where
\begin{eqnarray}
G_i := f_i - \fDef .
\end{eqnarray}

Starting from the ADM evolution equations, an evolution equation for
$f_i$ is obtained by differentiating (\ref{fEvolve}) and commuting
space and time derivatives.  The Hamiltonian and momentum constraints
are added with parameters $c$ and $b$, and derivatives of the
$G_i$ are added with parameters $a$ and $a'$:
\begin{eqnarray*}
\fl
\partial_t \gamma_{ij} & = & -2 \alpha K_{ij} \label{eqn:nor1}\\ \fl
\partial_t K_{ij} & = & -D_i D_j \alpha + \alpha(R^{(3)}_{ij} - 2
K_{ik} {K^k}_{j} + K_{ij} K) + \KadditionA + \KadditionB \\ \fl
\partial_t f_i & = & \alpha K^{kl} (2 \gamma_{ik,l} - \rho \gamma_{kl,i}) -
 \gamma^{kl} \left [ 2 (\alpha K_{ik}),_l - \rho (\alpha K_{kl}),_i \right ] + 2
b {\mathcal M}_i \\ \fl
\partial_t \alpha & = & -\alpha F(\alpha, K, x^i) .
\end{eqnarray*}
The variables $\gamma_{ij}$, $K_{ij}$, $f_i$ and $\alpha$ are evolved.
Due to the symmetries of $\gamma_{ij}$ and $K_{ij}$, this leads to 16
evolved variables. We write the Ricci tensor entirely in terms of
$\gamma_{ij}$; $f_i$ is only used where it appears as part of $G_i$.

For those tests requiring harmonic slicing, the lapse source function
is
\begin{eqnarray}
F(\alpha, K, x^i) = \alpha K
\end{eqnarray}
and for the expanding Gowdy test, 
\begin{eqnarray}\label{eq:HinderGowdyLapse}
F(\alpha, K, x^i) = K_{33}/\alpha
\end{eqnarray}
which is compatible with the exact lapse in this case.
We make the following choice of parameters:
\begin{eqnarray}
a = 1, \quad b = 1, \quad a' = 0, \quad \rho = 2/3, \quad c = 0 .
\end{eqnarray}
Note that choosing parameters
\begin{eqnarray}
a = 0, \quad b = 0, \quad a' = 0, \quad \rho = 0, \quad c = 0
\end{eqnarray}
leads to a standard ADM system. This is useful for testing the code.

%%%%%%%%%%%%%%%%%%%%%%%%%%%%%%%%%%%%%%%%%%%%%%%%%%
\subsubsection{Semi-discrete scheme:}
%%%%%%%%%%%%%%%%%%%%%%%%%%%%%%%%%%%%%%%%%%%%%%%%%%

To form the semi-discrete approximation, 
discretization in space is performed according
to the standard second order accurate discretization \ref{Eqstandiscr2}.

Finite differences are taken only of the evolved variables
$\gamma_{ij}$, $K_{ij}$, $f_i$ and $\alpha$. This means that where
derivatives of other quantities appear, they are explicitly written in
terms of derivatives of the evolved variables (e.g.\ by using the
Leibniz rule).  

We do not add Kreiss-Oliger type artificial dissipation, as it was not
necessary for stability.

%%%%%%%%%%%%%%%%%%%%%%%%%%%%%%%%%%%%%%%%%%%%%%%%%%
\subsubsection{Time integration:}
%%%%%%%%%%%%%%%%%%%%%%%%%%%%%%%%%%%%%%%%%%%%%%%%%%

Time integration is performed using the method of lines with the
iterative Crank-Nicholson (ICN) method.

%%%%%%%%%%%%%%%%%%%%%%%%%%%%%%%%%%%%%%%%%%%%%%%%%%
\subsubsection{Output:}
%%%%%%%%%%%%%%%%%%%%%%%%%%%%%%%%%%%%%%%%%%%%%%%%%%

For our state vector $v = ( \gamma_{ij}, K_{ij}, f_i)^T$ we
define the $L_2$ and $D_+$ norms:
\begin{eqnarray}
\|v\|_{L_2}^2 &\equiv& \sum_{\mbox{grid}}(\eta^{ik} \eta^{jl} \gamma_{ij} \gamma_{kl} 
   + \eta^{ik} \eta^{jl} K_{ij} K_{kl} + \eta^{ij} f_i f_j) h^3 \\
\|v\|_{D_+}^2 &\equiv& \|v\|_{L_2}^2 + \sum_{\mbox{grid}} 
 (\eta^{ik} \eta^{jl} \eta^{mn} D_{+m} \gamma_{ij} D_{+n} \gamma_{kl}) h^3
\end{eqnarray}
where $\eta \equiv \mbox{diag}(1,1,1)$.  This is the norm obtained
from a reduction to first order \cite{Calabrese:2005ft} of the semi-discrete
equations.  The exact solution is denoted $u^n_j \equiv u(t^n, x_j)$
and the error is defined as
\begin{eqnarray}
  {\cal E} \equiv v - u.
\end{eqnarray}
For the stability test, the exact solution is taken to be Minkowski in
Cartesian coordinates.  For those tests which are perturbations of
this solution, we analyze relative error with respect to this
background.  We denote the background solution as $u_B$. Hence the
relative error about this background is
\begin{eqnarray}
r \equiv \frac{\|{\cal E}\|_{L_2}}{\|u - u_B\|_{L_2}} .
\end{eqnarray}
In general, we run until this quantity exceeds 0.2 (a relative error
of 20\%).

%%%%%%%%%%%%%%%%%%%%%%%%%%%%%%%%%%%%%%%%%%%%%%%%%%%%%%%%%%%%%%%%%%%%%%
\subsection{Family of BSSN (Shibata-Nakamura and Baumgarte-Shapiro)
formulations}
%%%%%%%%%%%%%%%%%%%%%%%%%%%%%%%%%%%%%%%%%%%%%%%%%%%%%%%%%%%%%%%%%%%%%%

The family of BSSN systems is constituted by  variations of an evolution system
that had originally been proposed by Nakamura in the late 80s, and has been 
subsequently modified by Nakamura-Oohara and Shibata-Nakamura
\cite{Nakamura87,Nakamura89,Shibata95}, and later by various other authors. The
formulation is characterized by introducing a contracted connection term as a
new variable, a conformal decomposition of the metric and extrinsic curvature
variables, and adding constraints to the evolution equations. In particular, the
system can be viewed as the NOR-system plus a conformal decomposition which
leads to  the evolution of  a unimodular metric. The advantage of this
formulation was re-announced by Baumgarte and Shapiro \cite{Baumgarte99}.

Modifications of the system have been obtained by variations in how derivatives
of the new variables are written, how the gauge is specified, how algebraic
constraints are treated, and the way (differential or algebraic) constraints are
added to the evolution equations. A detailed discussion of well-posedness for
the BSSN family has been given by Gundlach and Martin-Garcia
\cite{Gundlach04a,Gundlach:2004jp,Gundlach:2005ta}, to which we refer for
details about the BSSN family.

The set of evolved variables are the logarithm of the conformal factor
$\varphi$, the conformally rescaled three-metric $\tilde \gamma_{ij}$,
the trace of the extrinsic curvature $K$, the conformally rescaled
traceless extrinsic curvature $\tilde A_{ij}$, and the contracted
Christoffel symbols $\tilde \Gamma^i$:

\begin{eqnarray}
\varphi &=& (1/12)\log ({\rm det}\gamma_{ij}),
\label{BSSNval1}
\\
\tilde{\gamma}_{ij} & = & e^{-4\varphi}\gamma_{ij},
\label{BSSNval2}
\\
K  &=& \gamma^{ij}K_{ij}, \label{BSSNval3}
\\
\tilde{A}_{ij} &=&
e^{-4\varphi}(K_{ij} - (1/3)\gamma_{ij}K), \label{BSSNval4}
\\
\tilde{\Gamma}^i &=&
\tilde{\Gamma}^i_{jk}\tilde{\gamma}^{jk} .
\label{BSSNval5}
\end{eqnarray}
This immediately leads to the two algebraic constraints
\begin{equation}
   \det \gamma_{ij} = 1, \qquad
   \tilde A^i_i     = 0  
\end{equation}
and the differential constraint
\begin{equation}
\tilde{\Gamma}^i-\tilde{\gamma}^{jk}\tilde{\Gamma}^i_{jk} 
= 0 \label{BSSNGammaconstraint},
\end{equation}
which are9 propagated by the evolution equations.
Note that densitized quantities (those with a tilde) have their indices raised
and lowered with the conformally rescaled three-metric $\tilde \gamma_{ij}$.

The standard Hamiltonian and momentum constraints (\ref{admCH},\ref{admCM}) and
(\ref{admCM}) can be expressed in the BSSN variables as
\begin{eqnarray}
{\cal H} &=&
e^{-4\varphi}\tilde{R}
-8e^{-4\varphi}\tilde{D}^j\tilde{D}_j\varphi
-8e^{-4\varphi}(\tilde{D}^j\varphi)(\tilde{D}_j\varphi)
+(2/3)K^2   \nonumber \\
 && -\tilde{A}_{ij}\tilde{A}^{ij}
-(2/3) {\cal A} K,
\label{BSSNconstraint1}
\\ 
{\cal M}_i
&=&
 6\tilde{A}^j{}_{i}(\tilde{D}_j \varphi)
-2{\cal A}(\tilde{D}_i \varphi)
-(2/3) (\tilde{D}_i K)
+\tilde{\gamma}^{kj}(\tilde{D}_j\tilde{A}_{ki}).
\label{BSSNconstraint2}
\end{eqnarray}

The BSSN evolution equations, which are obtained from the ADM equations
(\ref{admevo1} - \ref{admCM}) by using the definitions (\ref{BSSNval1} - \ref{BSSNval5})
and making a standard choice for adding constraints, are 
\begin{eqnarray}
\fl
{\cal L}_{n}\varphi &=& -(1/6)\alpha K, \label{BSSNeqmPHI} \\ \fl
{\cal L}_{n}\tilde{\gamma}_{ij} &=& -2\alpha\tilde{A}_{ij},\label{BSSNeqmtgamma}\\ \fl
{\cal L}_{n} K &=& -D^iD_i\alpha
+\alpha \tilde{A}_{ij}\tilde{A}^{ij} + (1/3) \alpha K^2, \label{BSSNeqmK} \\ \fl
{\cal L}_{n} \tilde{A}_{ij} &=&
-e^{-4\varphi}(D_iD_j\alpha)^{TF}
+e^{-4\varphi} \alpha (R^{BSSN}_{ij})^{TF}
+\alpha K\tilde{A}_{ij}
-2\alpha \tilde{A}_{ik}\tilde{A}^k{}_j, \label{BSSNeqmTA} \\ \fl
{\cal L}_{n} \tilde{\Gamma}^i &=& -2(\partial_j\alpha)\tilde{A}^{ij} +2\alpha
\big(\tilde{\Gamma}^i_{jk}\tilde{A}^{kj} -(2/3)\tilde{\gamma}^{ij}(\partial_j K)
+6\tilde{A}^{ij}(\partial_j\varphi) \big), \label{BSSNeqmTG}
\end{eqnarray}
where $\tilde{D}_i$ is the covariant derivative associated
with $\tilde{\gamma}_{ij}$, and
 ${\cal L}_n = \partial_t - {\cal L}_\beta$ is the Lie derivative along the
unit normal.
Note that $\int {\cal L}_{n} K d^3 x$ is positive definite apart
from boundary terms involving the lapse (which vanish for periodic
boundary conditions).
The Ricci curvature $R^{BSSN}_{ij}$ in terms of the BSSN variables becomes
\begin{eqnarray*} \fl
R^{BSSN}_{ij}
&=&
\tilde R_{ij}+R^\varphi_{ij},
\label{BSricci}
\\ \fl
R^\varphi_{ij}
&=&
-2\tilde{D}_i\tilde{D}_j\varphi
-2\tilde{\gamma}_{ij}\tilde{D}^k\tilde{D}_k\varphi
+4(\tilde{D}_i\varphi)(\tilde{D}_j\varphi)
-4\tilde{\gamma}_{ij}(\tilde{D}^k\varphi)(\tilde{D}_k\varphi),
\\ \fl
\tilde{R}_{ij}
&=&
-(1/2)\tilde{\gamma}^{lk}\partial_{l}\partial_{k}\tilde{\gamma}_{ij}
+\tilde{\gamma}_{k(i}\partial_{j)}\tilde{\Gamma}^k
+\tilde{\Gamma}^k\tilde{\Gamma}_{(ij)k}
+2\tilde{\gamma}^{lm}\tilde{\Gamma}^k_{l(i}\tilde{\Gamma}_{j)km}
+\tilde{\gamma}^{lm}\tilde{\Gamma}^k_{im}\tilde{\Gamma}_{klj}.
\end{eqnarray*}
Note that there are different ways to numerically compute the trace free
part of the Ricci tensor, e.g.\ one can project out the trace of the
Ricci tensor according to
\begin{equation}\label{eq:projectRicciTF}
R_{ij}^{TF} = R_{ij} - \frac{1}{3} R \gamma_{ij},
\end{equation}
compute the Ricci Scalar from the Hamiltonian constraint
(\ref{BSSNconstraint1}), or compute the trace free part explicitly by assuming
the algebraic constraints hold.

We refer to the code descriptions below for
details concerning the individual codes.
In summary, the fundamental dynamical variables in BSSN are
($\varphi,\tilde{\gamma}_{ij}$,
$K$,$\tilde{A}_{ij}$,$\tilde{\Gamma}^i$),
which total 17.
The 4 gauge quantities are ($\alpha, \beta^i$).

%%%%%%%%%%%%%%%%%%%%%%%%%%%%%%%%%%%%%%%%%%%%%%%%%%
\subsubsection{Concrete implementations}
%%%%%%%%%%%%%%%%%%%%%%%%%%%%%%%%%%%%%%%%%%%%%%%%%%

We have compared a number of codes based on variants of the BSSN system. Several
of these are based on the Cactus computational toolkit {\cite{Cactusweb}}: the
CCATIE\_BSSN \cite{Alcubierre99d, Alcubierre02a}
and Kranc\_BSSN \cite{FlexBSSN} codes, and the 
LazEv\_BSSN \cite{Zlochower2005:fourth-order} code. Of these,
CCATIE\_BSSN and Kranc\_BSSN use
the CactusMoL time integrator, which provides the RK3, RK4 and ICN methods,
among others  (see e.g.\ \cite{Calabrese:2005ft}). Kranc\_BSSN is based on the 
Kranc code generation software package \cite{Husa:2004ip}.

All codes use straightforward replacement of partial derivatives by standard
second order centered finite differences with a three point stencil
(most codes are also able to use standard centered fourth order finite
differencing).

Most of the BSSN codes have a long history of use in production environments and
have a large number of parameters that allow them great flexibility, e.g.\
regarding details of the numerical methods, gauge conditions, or the way the
algebraic constraints are treated. Typical options to solve the algebraic
constraints at every intermediate timestep use the following replacements:
\begin{itemize}
\item Ensure that $\tilde{\gamma}_{ij}$ has unit determinant by setting
\begin{equation}\label{eq:resetMetric}
\tilde{\gamma}_{ij} \to \frac {{\tilde\gamma}_{ij}} {\det {\tilde\gamma}^{1/3}}.
\end{equation}

\item Ensure that ${\tilde A}_{ij}$ remains trace-free by setting
\begin{equation}\label{eq:resetAij}
\tilde A_{ij} \to \tilde A_{ij} - \frac 1 3 \tilde A_{lm} \tilde \gamma ^{il}
\tilde \gamma^{jm}.
\end{equation}

\item Divide $\tilde A_{ij}$ by the same factor that is used to remove
the determinant of $\tilde \gamma_{ij}$:
\begin{equation}\label{eq:rescaleAij}
\tilde A_{ij} \to \frac {\tilde A_{ij}} {\det {\tilde\gamma}^{1/3}}.
\end{equation}
\end{itemize}

Note that an ambiguity arises whenever $\Gamma^i$ or $\tilde
\gamma^{kj} \gamma_{ij,k}$ occur, as they are related analytically by
the equation $\Gamma^i = - \gamma^{ij},_j - \frac 1 2 \gamma^{il} (\ln
\gamma),_l$.  If the constraint $\gamma = 1$ holds, e.g.\ if 
it is enforced at each timestep, this is equivalent numerically 
(up to round-off error) to $\Gamma^i = - \gamma^{ij},_j$.  
Some authors replace $\gamma^{ij},_j$ using
$-\Gamma^i$ only when the expression appears under a derivative, but more
complicated rules have also been applied.

Ref.~\cite{Brown2007b} describes a widely used combination of BSSN
system and gauge condition in detail and examines this system's
hyperbolicity.

%%%%%%%%%%%%%%%%%%%%%%%%%%%%%%%%%%%%%%%%%%%%%%%%%%
\subsection{KrancFN}\label{codeAEIKrancFN}
%%%%%%%%%%%%%%%%%%%%%%%%%%%%%%%%%%%%%%%%%%%%%%%%%%

%%%%%%%%%%%%%%%%%%%%%%%%%%%%%%%%%%%%%%%%%%%%%%%%%%
\subsubsection{Continuum formulation:}
%%%%%%%%%%%%%%%%%%%%%%%%%%%%%%%%%%%%%%%%%%%%%%%%%%

The Friedrich-Nagy system \cite{Friedrich99} is a frame-based first
order formulation that has been shown to yield a well-posed initial 
boundary value problem. The formulation starts from 
the four dimensional vacuum equations
\begin{eqnarray}
T_{IJ}{}^{\mu} &:= [e_I, e_J]^{\mu} - 
    (\Gamma_I{}^{K}{}_{J} - \Gamma_J{}^{K}{}_{I}) e_K{}^{\mu} 
             = 0, & \quad \mu = 0,1,2,3
\label{eq::torsionfree}\\
\Delta_{IJKL} &:= R_{IJKL}(\Gamma) - C_{IJKL} = 0 & 
\label{eq::vacuumEE} \\
H_{JKL} & := \nabla_I C_{JKL}{}^I  = 0,  &  \quad I=0,1,2,3 
\label{eq::Bianchi}
\end{eqnarray}
where $e_I$ denote the tetrad vectors with coordinate components
$e_I{}^{\mu}$; and $\Conn{I}{K}{J}$ are the connection coefficients
defined by $\nabla_{e_I} e_K = \Conn{I}{J}{K} e_J$ and satisfying
$\eta_{JM} \, \Conn{I}{J}{K}  + \eta_{KJ} \, \Conn{I}{J}{M} = 0$.
$R_{IJKL}$ and $C_{IJKL}$ denote the components of the 
Riemann and Weyl tensor with respect to the tetrad.
The Riemann tensor is given in terms of the connection coefficients 
by
\begin{eqnarray}
\fl   R_{IJ}{}^{L}{}_K(\Gamma)=e_I (\Conn{J}{L}{K})- e_J(\Conn{I}{L}{K}) 
         \nonumber \\
    - 
    \Conn{M}{L}{K} \Conn{I}{M}{J} -
    \Conn{I}{M}{K} \Conn{J}{L}{M} + 
    \Conn{M}{L}{K} \Conn{J}{M}{I} + 
    \Conn{I}{L}{M} \Conn{J}{M}{K}.
\end{eqnarray}
Equation (\ref{eq::torsionfree}) states that the connection is torsion free,
(\ref{eq::vacuumEE}) are the vacuum Einstein equations and 
(\ref{eq::Bianchi}) is the Bianchi identity for a vacuum spacetime.
From (\ref{eq::torsionfree}) -- (\ref{eq::Bianchi}), 
a symmetric hyperbolic evolution system is obtained by choosing certain 
combinations of components of the above equations as well as a gauge 
that is adapted to the boundary. 

Assuming a boundary at $z = const$, we foliate the interior domain by
time-like hypersurfaces $T_c$ given by $z = c = const$. The frame
is adapted to this foliation and boundary such that the
frame vector $e_3$ is orthogonal to $T_c$, which implies for the coordinate
components  
\begin{equation}\label{eq::ea3}
e_a{}^3 = 0, \quad a = 0,1,2, \quad e_3{}^{3} > 0.
\end{equation}
$e_3$ being the unit normal to $T_c$ implies 
$\Gamma_{a}{}^3{}_{b} = \Gamma_{(a}{}^3{}_{b)}$.

The mean extrinsic curvature of $T_c$ is prescribed as a 
function of the coordinates $f(x^{\mu})$ and used to eliminate the
connection coefficient $\Conn{0}{3}{0}$ from the equations,
\begin{equation}
\Gamma_0{}^3{}_0 = f + 
\Gamma_1{}^3{}_1 + \Gamma_2{}^3{}_2.
\end{equation}
The variation of $e_0$ within $T_c$ is prescribed by 
functions $F^A(x^{\mu})$, $A=1,2$ according to
$D_{e_0} e_0 = F^A e_A$, where $D$ denotes the induced connection on $T_c$.
This eliminates the connection coefficients
\begin{equation}\label{eq::aA}
\Conn{0}{A}{0} = F^A, \quad {A = 1,2}.
\end{equation}
The tetrad vectors $e_A$ are Fermi-transported along $e_0$
with respect to $D$ and therefore
\begin{equation}
\Conn{0}{A}{B} = 0, \quad {A,B = 1,2} .
\end{equation}
The coordinates $\{x^{\mu}\}$
are chosen such that the tetrad vector $e_0$ represents the time flow 
$\partial_t$, i.e.,
\begin{equation}\label{eq::timeflow}
e_0{}^{\mu} = \delta_0{}^{\mu}.
\end{equation}

The ten independent components of the Weyl tensor are encoded in the 
symmetric and tracefree tensor fields
$$
E_{ij} := C_{i0j0}, \qquad
B_{ij} := \frac{1}{2} C_{0 i k l} {}^{\scriptscriptstyle(3)}\!\epsilon^{kl}{}_j
$$
corresponding to the electric and magnetic parts with respect to $e_0$.
The conditions $\delta^{ij} E_{ij} = \delta^{ij} B_{ij} = 0$ are
incorporated explicitly by eliminating 
\begin{equation}\label{eq::WeylTF}
E_{33} = - (E_{11} + E_{22}), \quad  B_{33} = - (B_{11} + B_{22})
\end{equation}
from the equations.
In total the Friedrich-Nagy system has 37 variables, namely
\begin{equation}\label{eq::varsexplicit}
{\bf u} = (e_A{}^{p}, e_3{}^{\mu}, \Conn{i}{0}{j},
\Conn{3}{i}{j},\Conn{(A}{3}{B)}, \Conn{A}{B}{C}, E_{iA}, B_{iA})^T,
\end{equation}
where
$$
{A,B,C=1,2, \quad i, j = 1,2,3, \quad p = 0,1,2, \quad \mu = 0,1,2,3}.
$$

A symmetric hyperbolic evolution system for the variables
(\ref{eq::varsexplicit}) is obtained by taking the following 
combinations of (\ref{eq::torsionfree}) -- (\ref{eq::Bianchi}):
\begin{eqnarray*}
\fl
& & T_{0A}{}^{p} = 0, \quad T_{03}{}^{\mu} = 0, \quad \Delta_{0 B a b} = 0,  \quad \Delta_{0131} = 0,
    \quad \Delta_{0232} = 0, \\ \fl
& & \Delta_{0132}+\Delta_{0231} = 0, \quad \Delta_{0130} + \Delta_{1232} = 0, 
    \quad \Delta_{0230}+\Delta_{2131} = 0, \\ \fl
& & \Delta_{AB03} = 0, \quad \Delta_{A003} = 0, 
     \quad \Delta_{3A03} + \Delta_{303A} = 0, \quad 
 \eta^{ab} \Delta_{3ab3} = 0, \\ \fl
& & H_{0ij} - \frac{1}{2} \delta^3{}_{(i} \epsilon_{j)}{}^{3l}  H_{mn0}
    \, \epsilon^{mn}{}_l = 0, \quad  
  \frac{1}{2} H_{mki} \, \epsilon^{mk}{}_j + \delta^3{}_{(i}
\epsilon_{j)}{}^{3 m} \, H_{0 m0} = 0
\end{eqnarray*}
where the convention for the indices is the same as in
Eq.~(\ref{eq::varsexplicit}) and $a,b = 0,1,2$. 
The resulting system is given explicitly in \cite{Friedrich99, Alic05}
and is of the form
\begin{equation}\label{eq::FullEqs}
{\bf A}^0 \partial_t {\bf u} + {\bf A}^i \partial_i {\bf u}  
+ {\bf B}({\bf u}, F) = 0,
\end{equation}
where $F = (f, F^A, \partial_{\mu} f, \partial_{\mu} F^A)$ represents the gauge
source functions and their derivatives.  The matrices ${\bf A}^0, {\bf A}^i$ are
symmetric and depend on the coordinate components of the frame. ${\bf A}^0$ is
positive definite  as long as $1 - (e_1{}^0)^2 - (e_2{}^0)^2 - (e_3{}^0)^2 >
0$,  which corresponds to $e_0$ being time-like. Characteristics are time-like
and null.

The remaining components of (\ref{eq::torsionfree})--(\ref{eq::Bianchi}), 
$$
     T_{ij}{}^{\mu} = 0, \quad
    \Delta_{ij}{}^L{}_K = 0, \quad 
      H_{0k0} = 0, \quad \frac{1}{2} H_{jk0} \epsilon^{jk}{}_m = 0,
$$
only contain derivatives in directions orthogonal to $e_0$ and are satisfied
if satisfied initially by virtue of the evolution equations 
(see \cite{Friedrich99}). 
$e_0$ in general is not hypersurface orthogonal and therefore 
the constraints do contain derivatives in direction of $\partial_t$.
In order to monitor these constraints during a numerical evolution, 
we eliminate the time derivatives by means of the evolution equations.

%%%%%%%%%%%%%%%%%%%%%%%%%%%%%%%%%%%%%%%%%%%%%%%%%%
\subsubsection{Numerical implementation:}
%%%%%%%%%%%%%%%%%%%%%%%%%%%%%%%%%%%%%%%%%%%%%%%%%%
The code is based on the Cactus Computational Toolkit \cite{Cactusweb} and
the Kranc software \cite{Husa:2004ip,Alic05}.
The spatial discretization of (\ref{eq::FullEqs}) is done in
a straight forward way
\begin{equation}\label{FNdiscr}
\partial_t {\bf u} = - ({\bf A}^0)^{-1} {\bf A}^i D_i {\bf u} + 
      ({\bf A}^0)^{-1}{\bf B}({\bf u}, F), 
\end{equation}
where $D_i$ is the 2nd (or 4th order) accurate 
centered derivative operator in the direction $i$ 
(see \ref{sec:appendix_FD}). Time integration is done with the method of
lines (CactusMoL) using ICN for the 2nd order scheme and RK4 for the 
4th order scheme.
If needed, artificial dissipation is added to the right hand side of 
equation (\ref{FNdiscr}) in the form
\begin{equation}
({\bf A}^0)^{-1} Q_d \, {\bf u},
\end{equation}
where $Q_d$ is the Kreiss-Oliger dissipation operator 
(see \ref{sec:appendix_dissipation}). Respecting the symmetrizer in the
dissipation term is essential; replacing it by the identity matrix triggered
exponentially growing continuum modes e.g.\ for the gauge wave testbed
with non-linear amplitude.

%%%%%%%%%%%%%%%%%%%%%%%%%%%%%%%%%%%%%%%%%%%%%%%%%%%%%%%%%%%%%%%%%%%%%%%%%%%%
\subsection{LSU\_HyperGR}
%%%%%%%%%%%%%%%%%%%%%%%%%%%%%%%%%%%%%%%%%%%%%%%%%%%%%%%%%%%%%%%%%%%%%%%%%%%%

This symmetric hyperbolic first order formulation is described by Sarbach and
Tiglio in \cite{Sarbach02b}. The system has 34 evolved variables which are
the standard ADM metric $\gamma_{i j}$, extrinsic curvature $K_{i j}$ 
and lapse $\alpha$, as well as extra variables
$d_{k i j} = \partial_k \gamma_{i j}$ and 
$A_i = \partial_i \alpha / \alpha$, introduced to make the formulation
first order in space.

In addition to the Hamiltonian constraint $\mathcal{H}$
and the momentum constraint $\mathcal{M}_i$, 
the constraints arising from those new variables are
\begin{eqnarray}
  C_{A_i} &=& A_i - \partial_i \alpha / \alpha, \\
  C_{k i j} &=& d_{k i j} - \partial_k \gamma_{i j}, \\
  C_{l k i j} &=& \partial_{[l} d_{k] j k}.
\end{eqnarray}

The system of PDEs resulting from the standard ADM 3+1 decomposition of the
Einstein equations is only weakly hyperbolic. To get a symmetric hyperbolic
system the principal part has to be modified further. This is done by adding the
constraints to the right hand sides of the evolution equations with appropriate 
multiplicative factors $\zeta, \xi, \eta, \chi$ and $\iota$.  Here these
parameters are chosen to be constant in space, although in general this is not
necessary. The full set of equations  is then
\begin{eqnarray}
\fl  \partial_0 \gamma_{i j} &=& -2 K_{i j}, \\ \fl
  \partial_0 K_{i j} &=& R_{i j} - \frac{1}{\alpha} \nabla_i \nabla_j \alpha 
           - 2 K_{i a} K^a_{\ j} + K K_{i j} + \iota \gamma_{i j} \mathcal{H} 
           + \zeta \gamma^{a b} C_{a (i j) b}, \\ \fl
  \partial_0 d_{k i j} &=& -2 \partial_k K_{i j} - 2 A_k K_{i j} +
   \eta \gamma_{k (i} \mathcal{M}_{j)} + \chi \gamma_{i j} \mathcal{M}_k, \\ \fl
  \partial_0 \alpha &=& -F(\alpha,K,x^\mu) + S(x^\mu), \\ \fl
  \partial_0 A_i &=& -{\partial F(\alpha,K,x^\mu) \over \partial \alpha} A_i -
    {1 \over \alpha} {\partial F(\alpha,K,x^\mu) \over \partial K} \partial_i K
    - {1 \over \alpha} {\partial F(\alpha,K,x^\mu) \over \partial x^i} + 
    \xi \mathcal{M}_i,
\end{eqnarray}
where $\partial_0 = (\partial_t - {\cal L}_\beta) / \alpha$,  $R_{i j}$ is the
Ricci tensor  and  $K$  the trace of the extrinsic curvature. The functions
$F(\alpha,K,x^i)$ and $S(x^i)$ are pure gauge and can be chosen freely. The
choices $S=0$ and $F=\alpha K$ provides harmonic gauge conditions.

Restriction of the parameters ${\chi, \xi, \eta, \zeta, \iota}$ to the 
family
\begin{equation}
  \iota = -1/2,{\rm \ } \zeta \eta = 
  -2,{\rm \ } \xi = -1/2 \chi + 1/4 \eta - 1/2
\end{equation}
results in a strongly hyperbolic system. A symmetric hyperbolic subfamily is
given by  $\zeta = -1$, which leaves $\chi$ as the single
free parameter (constrained only by the condition
$\chi \neq 0$). The runs presented here were done with the specific choice of $\chi=-1$.

To ensure a numerically stable discretization based on the energy method
for hyperbolic equations, second order spatial differencing operators 
that satisfy the  summation by parts (SBP) condition are used
\cite{Strand1994a, Lehner2005a}.

Furthermore  a small amount of dissipation
(standard Kreiss-Oliger dissipation operators)
is added to the right hand sides of the evolution equations.

The integration in time is done with a third order Runge-Kutta scheme.

%%%%%%%%%%%%%%%%%%%%%%%%%%%%%%%%%%%%%%%%%%%%%%%%%%
\section{Numerical methods}\label{sec:appendix_numerics}
%%%%%%%%%%%%%%%%%%%%%%%%%%%%%%%%%%%%%%%%%%%%%%%%%%

%%%%%%%%%%%%%%%%%%%%%%%%%%%%%%%%%%%%%%%%%%%%%%%%%%
\subsection{Spatial discretization}\label{sec:appendix_FD}
%%%%%%%%%%%%%%%%%%%%%%%%%%%%%%%%%%%%%%%%%%%%%%%%%%

Most of our numerical results are based on {\em second order
 accurate centered discretization}:
\begin{eqnarray}
&&\partial_i \to D_{0i}\,,\qquad \partial_i\partial_j \to
\left\{ \begin{array}{ll}D_{0i}D_{0j} & \mbox{if $i \neq j$}\\
D_{+i}D_{-i} & \mbox{if $i =j$}\end{array}\right.\,,
\label{Eqstandiscr2}
\end{eqnarray}
where 
\begin{eqnarray}
      D_+v_j &:=& \frac{v_{j+1}-v_j}{\Delta x}, \\
      D_-v_j &:=& \frac{v_{j}-v_{j-1}}{\Delta x}, \nonumber \\
      D_0v_j &:=& \frac{v_{j+1}-v_{j-1}}{2\Delta x}, \nonumber \\
      D_+D_-v_j &:=& \frac{v_{j+1}-2v_j+v_{j-1}}{\Delta x^2}.
\end{eqnarray}
For a summary of definitions and results for standard
fourth order discretizations we again refer to \cite{Calabrese:2005ft},
where in particular some results concerning the evolution systems 
considered here are derived.

Finally, averaging operators $A_{\pm}$  are defined as:
\begin{eqnarray}
A_{+} v_j := \frac{v_{j+1} + v_{j}}{2} \\
A_{-} v_j := \frac{v_{j} + v_{j-1}}{2} .
\end{eqnarray}

%%%%%%%%%%%%%%%%%%%%%%%%%%%%%%%%%%%%%%%%%%%%%%%%%%
\subsection{Artificial Dissipation}\label{sec:appendix_dissipation}
%%%%%%%%%%%%%%%%%%%%%%%%%%%%%%%%%%%%%%%%%%%%%%%%%%

For second order accurate codes, it is common practice to add 
third order accurate 
Kreiss--Oliger dissipation \cite{Kreiss73} to all right-hand-sides of
the time evolution equations as
\begin{equation}
\partial_t {\bf u} \rightarrow \partial_t {\bf u} + Q {\bf u}.
\end{equation}
Here we use the following
general form of the Kreiss--Oliger dissipation
operator $Q$ of order $2r$,
\begin{equation}\label{eq:KOdissipation}
Q = \sigma (-1)^r h^{2 r - 1} (D_+)^{r} \rho (D_-)^{r}/2^{2r},
\end{equation}
for a $2r -2$ accurate scheme, where the parameter $\sigma$ regulates the
strength of the dissipation and $\rho$ is a  weighting function, which is
typically set to 1 in the interior but may go to 0 at a boundary. Since we
mostly focus on second order accurate codes here, the relevant case is $r=2$,
for which \begin{equation} Q = -\sigma h^{3} (D_+)^{2} \rho (D_-)^{2}/16,
\end{equation} which may be implemented using Erik Schnetter's  Cactus thorn
{\tt AEIThorns/Dissipation} \cite{AEIThorns-Dissipation}.

%%%%%%%%%%%%%%%%%%%%%%%%%%%%%%%%%%%%%%%%%%%%%%%%%%%%%%%%%%%%%%%%%%%%%%
%%%%%%%%%%%%%%%%%%%%%%%%%%%%%%%%%%%%%%%%%%%%%%%%%%%%%%%%%%%%%%%%%%%%%%
%%%%%%%%%%%%%%%%%%%%%%%%%%%%%%%%%%%%%%%%%%%%%%%%%%%%%%%%%%%%%%%%%%%%%%

%%%%%%%%%%%%%%%%%%%%%%%%%%%%%%%%%%%%%%%%%%%%%%%%%%
\section*{References}
%%%%%%%%%%%%%%%%%%%%%%%%%%%%%%%%%%%%%%%%%%%%%%%%%%

\bibliographystyle{unsrt}%
\bibliography{refs}

\begin{thebibliography}{10}

\bibitem{Hahn64}
Susan~G. Hahn and Richard~W. Lindquist.
\newblock The two body problem in geometrodynamics.
\newblock {\em Ann. Phys.}, 29:304--331, 1964.

\bibitem{Pretorius:2005gq}
Frans Pretorius.
\newblock Evolution of binary black hole spacetimes.
\newblock {\em Phys. Rev. Lett.}, 95:121101, 2005.

\bibitem{Campanelli:2005dd}
Manuela Campanelli, Carlos~O. Lousto, Pedro Marronetti, and Yosef Zlochower.
\newblock Accurate evolutions of orbiting black-hole binaries without excision.
\newblock {\em Phys. Rev. Lett.}, 96:111101, 2006.

\bibitem{Baker:2005vv}
John~G. Baker, Joan Centrella, Dae-Il Choi, Michael Koppitz, and James van
  Meter.
\newblock Gravitational wave extraction from an inspiraling configuration of
  merging black holes.
\newblock {\em Phys. Rev. Lett.}, 96:111102, 2006.

\bibitem{applesweb}
Apples {W}ith {A}pples: {N}umerical {R}elativity {C}omparisons and {T}ests:
  {\tt http://www.ApplesWithApples.org}.

\bibitem{Alcubierre2003:mexico-I}
Miguel Alcubierre, Gabrielle Allen, Thomas~W. Baumgarte, Carles Bona, David
  Fiske, Tom Goodale, Francisco~Siddhartha Guzm{\'a}n, Ian Hawke, Scott Hawley,
  Sascha Husa, Michael Koppitz, Christiane Lechner, Lee Lindblom, Denis
  Pollney, David Rideout, Marcelo Salgado, Erik Schnetter, Edward Seidel, Hisa
  aki Shinkai, Deirdre Shoemaker, B{\'e}la Szil{\'a}gyi, Ryoji Takahashi, and
  Jeffrey Winicour.
\newblock Towards standard testbeds for numerical relativity.
\newblock {\em Class. Quantum Grav.}, 21(2):589--613, 2004.

\bibitem{Beyer:2004sv}
Horst Beyer and Olivier Sarbach.
\newblock {O}n the well posedness of the {B}aumgarte-{S}hapiro-
  {S}hibata-{N}akamura formulation of {E}instein's field equations.
\newblock {\em Phys. Rev. D}, 70:104004, 2004.

\bibitem{Gundlach04a}
C.~Gundlach and J.M. Martin-Garcia.
\newblock Symmetric hyperbolic form of systems of second-order evolution
  equations subject to constraints.
\newblock {\em Phys. Rev. D}, 70:044031, 2004.

\bibitem{Gundlach:2004jp}
Carsten Gundlach and Jose~M. Martin-Garcia.
\newblock Symmetric hyperbolicity and consistent boundary conditions for
  second-order {E}instein equations.
\newblock {\em Phys. Rev. D}, 70:044032, 2004.

\bibitem{Gundlach:2005ta}
Carsten Gundlach and Jose~M. Martin-Garcia.
\newblock Hyperbolicity of second-order in space systems of evolution
  equations.
\newblock {\em Class. Quantum Grav.}, 23:S387--S404, 2006.

\bibitem{Calabrese:2005ft}
Gioel Calabrese, Ian Hinder, and Sascha Husa.
\newblock Numerical stability for finite difference approximations of
  {E}instein's equations.
\newblock {\em J. Comp. Phys.}, 218:607--634, 2006.

\bibitem{Szilagyi05}
B.~Szil{\'a}gyi, H-O. Kreiss, and J.~Winicour.
\newblock Modeling the black hole excision problem.
\newblock {\em Phys. Rev. D}, 71:104035, 2005.

\bibitem{Motamed06}
Mohammad Motamed, M.~C. Babiuc, B.~Szilagyi, H-O. Kreiss, and J.Winicour.
\newblock Finite difference schemes for second order systems describing black
  holes.
\newblock {\em Phys. Rev. D}, 73:124008, 2006.

\bibitem{cfd-online-web-tests}
{\it Computational Fluid Dynamics Wiki / Validation and test cases}, CFD
  Online, {\tt http://www.cfd-online.com/Wiki/Validation\_and\_test\_cases}.

\bibitem{ODEIVPweb}
Test set~for IVP~solvers.
\newblock \texttt{http://pitagora.dm.uniba.it/$\sim$testset/}.

\bibitem{Courant62b}
R.~Courant and K.~O. Friedrichs.
\newblock {\em Supersonic flows and shock waves}.
\newblock Springer, Berlin, 1976.

\bibitem{Nagy:2004td}
G.~Nagy, O.~E. Ortiz, and O.~A. Reula.
\newblock Strongly hyperbolic second order {E}instein's evolution equations.
\newblock {\em Phys. Rev. D}, 70:044012, 2004.

\bibitem{Gundlach:2006tw}
Carsten Gundlach and Jose~M. Martin-Garcia.
\newblock Well-posedness of formulations of the {E}instein equations with
  dynamical lapse and shift conditions.
\newblock {\em Phys. Rev. D}, 74:024016, 2006.

\bibitem{Calabrese02a}
G.~Calabrese, J.~Pullin, O.~Sarbach, and M.~Tiglio.
\newblock Convergence and stability in numerical relativity.
\newblock {\em Phys. Rev. D}, 66:041501, 2002.

\bibitem{Gustafsson95}
Bertil Gustafsson, Heinz-Otto Kreiss, and Joseph Oliger.
\newblock {\em Time dependent problems and difference methods}.
\newblock Wiley, New York, 1995.

\bibitem{Babiuc:2004pi}
Maria~C. Babiuc, B{\'e}la Szil{\'a}gyi, and J.~Winicour.
\newblock Some mathematical problems in numerical relativity.
\newblock {\em Lect. Notes Phys.}, 692:251--274, 2006.

\bibitem{Husa:2005ns}
Sascha Husa, Carsten Schneemann, Tilman Vogel, and Anil Zenginoglu.
\newblock Hyperboloidal data and evolution.
\newblock 2005.
\newblock To appear in Proceedings of the 2005 spanish relativity meeting, AIP
  Conference Proceedings, 8 pages.

\bibitem{Babiuc-etal-2005}
Maria~C. Babiuc, B{\'e}la Szil{\'a}gyi, and Jeffrey Winicour.
\newblock Testing numerical relativity with the shifted gauge wave.
\newblock {\em Class. Quantum Grav.}, 23:S319--S342, 2006.

\bibitem{Babiuc:2006wk}
Maria~C. Babiuc, B{\'e}la Szil{\'a}gyi, and J.Winicour.
\newblock Harmonic initial-boundary evolution in general relativity.
\newblock {\em Phys. Rev. D}, 73:064017, 2006.

\bibitem{Boyle07}
M.~Boyle, L.~Lindblom, H.~Pfeiffer, M.~Scheel, and L.~Kidder.
\newblock Testing the accuracy and stability of spectral methods in numerical
  relativity.
\newblock {\em Phys. Rev.}, D75:024006--024018, 2007.

\bibitem{Kidder01a}
L.~E. Kidder, Mark~A. Scheel, and Saul~A. Teukolsky.
\newblock Extending the lifetime of 3{D} black hole computations with a new
  hyperbolic system of evolution equations.
\newblock {\em Phys. Rev. D}, 64:064017, 2001.

\bibitem{Gowdy71}
R.~H. Gowdy.
\newblock {\em Phys. Rev. Lett.}, 27:826, 1971.

\bibitem{Ringstrom03}
H.~Ringstrom.
\newblock On a wave map equation arising in general relativity.
\newblock 2003.

\bibitem{Berger02}
B.K. Berger.
\newblock Asymptotic behavior of a class of expanding gowdy spacetimes.
\newblock {\em submitted to Phys. Rev. D}, 2002.

\bibitem{Frauendiener:2004bj}
Jorg Frauendiener and Tilman Vogel.
\newblock Algebraic stability analysis of constraint propagation.
\newblock {\em Class. Quantum Grav.}, 22:1769--1793, 2005.

\bibitem{Husa02b}
Sascha Husa.
\newblock In L.~Fern{\'a}ndez and L.~Manuel Gonz{\'a}lez, editors, {\em Current
  trends in relativistic astrophysics}, volume 617 of {\em Lecture Notes in
  Physics}. Springer, 2002.

\bibitem{Garfinkle02}
David Garfinkle.
\newblock Harmonic coordinate method for simulating generic singularities.
\newblock {\em Phys. Rev. D}, 65:044029, 2002.

\bibitem{Post05}
D.~E. Post and L.~G. Votta.
\newblock Computational science demands a new paradigm.
\newblock {\em Physics Today.}, 58:35, 2005.

\bibitem{Rinne06}
Oliver Rinne.
\newblock Stable radiation-controlling boundary conditions for the generalized
  harmonic einstein equations.
\newblock {\em Class. Quantum Grav.}, 23:6275--6300, 2006.

\bibitem{Husa:2004ip}
Sascha Husa, Ian Hinder, and Christiane Lechner.
\newblock Kranc: a {Mathematica} application to generate numerical codes for
  tensorial evolution equations.
\newblock {\em Comput. Phys. Comm.}, 174:983--1004, 2006.

\bibitem{Alic05}
D.~Alic.
\newblock Toward the numerical implementation of well-posed, constraint
  preserving evolution systems for general relativity.
\newblock Master's thesis, University of Timisoara, 2005.

\bibitem{Hinder2005}
Ian Hinder.
\newblock {\em Well-posed formulations and stable finite differencing schemes
  for numerical relativity}.
\newblock PhD thesis, University of Southampton, Southampton, UK, 2005.

\bibitem{AEIThorns-Dissipation}
Erik Schnetter.
\newblock {\tt AEIThorns/Dissipation} Cactus thorn.

\bibitem{Arnowitt62}
Richard Arnowitt, Stanley Deser, and Charles~W. Misner.
\newblock The dynamics of general relativity.
\newblock In L.~Witten, editor, {\em Gravitation: An introduction to current
  research}, pages 227--265. John Wiley, New York, 1962.

\bibitem{York79}
James~W. York.
\newblock Kinematics and dynamics of general relativity.
\newblock In Larry~L. Smarr, editor, {\em Sources of gravitational radiation},
  pages 83--126. Cambridge University Press, Cambridge, UK, 1979.

\bibitem{Frittelli97a}
Simonetta Frittelli.
\newblock Note on the propagation of the constraints in standard 3+1 general
  relativity.
\newblock {\em Phys. Rev. D}, 55:5992--5996, 1997.

\bibitem{Cactusweb}
Cactus~Computational Toolkit.
\newblock \texttt{http://www.cactuscode.org}.

\bibitem{Szilagyi02a}
B.~Szil{\'a}gyi and Jeffrey Winicour.
\newblock Well-posed initial-boundary evolution in general relativity.
\newblock {\em Phys. Rev. D}, 68:041501, 2003.

\bibitem{Nakamura87}
Takashi Nakamura, {Ken-ichi} Oohara, and Yasufumi Kojima.
\newblock General relativistic collapse to black holes and gravitational waves
  from black holes.
\newblock {\em Prog. Theor. Phys. Suppl.}, 90:1--218, 1987.

\bibitem{Nakamura89}
Takashi Nakamura and {Ken-ichi} Oohara.
\newblock Methods in {3D} numerical relativity.
\newblock In C.~Evans, L.~Finn, and D.~Hobill, editors, {\em Frontiers in
  Numerical Relativity}, pages 254--280. Cambridge University Press, Cambridge,
  England, 1989.

\bibitem{Shibata95}
Masaru Shibata and Takashi Nakamura.
\newblock Evolution of three-dimensional gravitational waves: {H}armonic
  slicing case.
\newblock {\em Phys. Rev. D}, 52:5428, 1995.

\bibitem{Baumgarte99}
Thomas~W. Baumgarte and Stuart~L. Shapiro.
\newblock On the numerical integration of {E}instein's field equations.
\newblock {\em Phys. Rev. D}, 59:024007, 1999.

\bibitem{Alcubierre99d}
Miguel Alcubierre, Bernd Br{\"u}gmann, Thomas Dramlitsch, Jos{\'e}~A. Font,
  Philippos Papadopoulos, Edward Seidel, Nikolaos Stergioulas, and Ryoji
  Takahashi.
\newblock Towards a stable numerical evolution of strongly gravitating systems
  in general relativity: The conformal treatments.
\newblock {\em Phys. Rev. D}, 62:044034, 2000.

\bibitem{Alcubierre02a}
Miguel Alcubierre, Bernd Br{\"u}gmann, Peter Diener, Michael Koppitz, Denis
  Pollney, Edward Seidel, and Ryoji Takahashi.
\newblock Gauge conditions for long-term numerical black hole evolutions
  without excision.
\newblock {\em Phys. Rev. D}, 67:084023, 2003.

\bibitem{FlexBSSN}
Sascha Husa.
\newblock The FlexBSSN code is available on request from the author.

\bibitem{Zlochower2005:fourth-order}
Y.~Zlochower, J.~G. Baker, M.~Campanelli, and C.~O. Lousto.
\newblock Accurate black hole evolutions by fourth-order numerical relativity.
\newblock {\em Phys. Rev. D}, 72:024021, 2005.

\bibitem{Brown2007b}
David Brown, Olivier Sarbach, Erik Schnetter, Manuel Tiglio, Peter Diener, Ian
  Hawke, and Denis Pollney.
\newblock Excision without excision.
\newblock {\em Phys. Rev. D}, 76:081503(R), 2007.

\bibitem{Friedrich99}
Helmut Friedrich and Gabriel Nagy.
\newblock The initial boundary value problem for {E}instein's vacuum field
  equations.
\newblock {\em Commun. Math. Phys.}, 201:619--655, 1999.

\bibitem{Sarbach02b}
O.~Sarbach and M.~Tiglio.
\newblock Exploiting gauge and constraint freedom in hyperbolic formulations of
  {E}instein's equations.
\newblock {\em Phys. Rev. D}, 66:064023, 2002.

\bibitem{Strand1994a}
B.~Strand.
\newblock Summation by parts for finite differencing approximations for d/dx.
\newblock {\em J. Comput. Phys.}, 110:47, 1994.

\bibitem{Lehner2005a}
Luis Lehner, Oscar Reula, and Manuel Tiglio.
\newblock Multi-block simulations in general relativity: high order
  discretizations, numerical stability, and applications.
\newblock {\em Class. Quantum Grav.}, 22:5283--5322, 2005.

\bibitem{Kreiss73}
Heinz~Otto Kreiss and Joseph Oliger.
\newblock {\em Methods for the approximate solution of time dependent
  problems}.
\newblock GARP publication series No. 10, Geneva, 1973.

\end{thebibliography}

\end{document}